%% file: main3.0.tex
\documentclass[aps,twocolumn,prl,superscriptaddress,amsmath,amssymb]{revtex4-1}
\newcommand{\bras}[1]{\langle#1\rvert}
\newcommand{\kets}[1]{\lvert#1\rangle}

\newcommand{\ket}[1]{\left|#1\right>}

\newcommand{\means}[1]{\langle#1\rangle}

\newcommand{\nematic}{\psi^{}_\text{N}}
\newcommand{\fidelity}{F^\text{sq}_\text{N}}
\newcommand{\ketbasic}[1]{\lvert{#1}\rangle}
\newcommand{\brabasic}[1]{\langle{#1}\rvert}
\newcommand{\braketbasic}[2]{\langle{#1}\vert{#2}\rangle}
\newcommand{\up}{\uparrow}
\newcommand{\Up}{\Uparrow}
\newcommand{\dw}{\downarrow}
\newcommand{\Dw}{\Downarrow}
\newcommand{\ra}{\rightarrow}
\newcommand{\Ra}{\Rightarrow}
\newcommand{\la}{\leftarrow}
\newcommand{\La}{\Leftarrow}

\ifx\pdfoutput\undefined
\usepackage[dvipdfmx]{graphicx}
\usepackage[dvipdfmx]{hyperref}
\else
\usepackage{graphicx}
\usepackage{hyperref}
\fi

\usepackage{txfonts}
\usepackage[T1]{fontenc}
\usepackage{xspace}

\usepackage{dcolumn}
\usepackage{bm}
\usepackage{amsmath}
\usepackage{booktabs}
\usepackage[usenames]{color}
\usepackage{xcolor}

\usepackage{ulem}

\begin{document}
\let\emph\textit

%\draft
%\preprint{?????}
\title{
Spin-Liquid--to--Spin-Liquid Transition in Kitaev Magnets Driven by Fractionalization
}

\author{Joji Nasu}
\affiliation{Department of Physics, Tokyo Institute of Technology, Meguro, Tokyo 152-8551, Japan}
\author{Yasuyuki Kato}
\affiliation{Department of Applied Physics, University of Tokyo, Bunkyo, Tokyo 113-8656, Japan}
\author{Junki Yoshitake}
\affiliation{Department of Applied Physics, University of Tokyo, Bunkyo, Tokyo 113-8656, Japan}
\author{Yoshitomo Kamiya}
\affiliation{Condensed Matter Theory Laboratory, RIKEN, Wako, Saitama 351-0198, Japan}
\author{Yukitoshi Motome}
\affiliation{Department of Applied Physics, University of Tokyo, Bunkyo, Tokyo 113-8656, Japan}

 \date{\today}
 \begin{abstract}
  While phase transitions between magnetic analogs of three states of matter --- a long-range ordered state, paramagnet, and spin liquid --- have been extensively studied, the possibility of ``liquid-liquid'' transitions, namely, between different spin liquids, remains elusive.
  By introducing the additional Ising coupling into the honeycomb Kitaev model with bond asymmetry, we discover that the Kitaev spin liquid turns into a spin-nematic quantum paramagnet before a magnetic order is established by the Ising coupling.
  The quantum phase transition between the two liquid states accompanies a topological change driven by fractionalized excitations, the $Z_2$ gauge fluxes, and is of first order.
  At finite temperatures, this yields a persisting first-order transition line that terminates at a critical point located deep inside the regime where quantum spins are fractionalized.
  It is suggested that similar transitions may occur in other perturbed Kitaev magnets with bond asymmetry.
 \end{abstract}

\maketitle

  Quantum spin liquids (QSLs) have been the subject of great interest since Anderson's inspiring proposal~\cite{Anderson1973153}.
  In QSLs, localized spins do not solidify into a long-range magnetic order even at zero temperature ($T$) despite strong interactions, analogous to a quantum fluid state of liquid helium.
  The intensively discussed candidates include some organic salts~\cite{PhysRevLett.91.107001,ISI:000278318600025,PhysRevLett.112.177201} and transition metal compounds~\cite{ISI:000231836700036,PhysRevLett.98.107204,PhysRevLett.99.137207} with geometric frustration.
  QSLs have brought rich physics as done by liquid helium.
  For instance, topological orders need to be introduced to characterize gapped QSLs~\cite{Wen2004,Lacroix2011C16}, in which fractionalization of quantum spins yields excitations with emergent statistics~\cite{PhysRevLett.96.060601}.
  For certain gapless QSLs, emergent fermionic excitations can be regarded as a hallmark, which has motivated to investigate asymptotic low-$T$ behavior of the specific heat and thermal conductivity~\cite{yamashita2008thermodynamic,yamashita2009thermal}.

  In his seminal work~\cite{Kitaev2006}, Kitaev proposed the canonical model of QSLs.
  This simple $S=1/2$ model with bond-dependent interactions provides exact realizations of QSLs with topological order and fractional excitations~\cite{Kitaev2006}.
  By introducing Majorana fermions, it can be shown that the ground state is a $Z_2$ QSL either gapped or gapless depending on the exchange parameters~\cite{PhysRevLett.98.247201}.
  Moreover, exchange interactions in some transition metal compounds with strong spin-orbit coupling may be dominated by the Kitaev-type ones~\cite{PhysRevLett.102.017205}.
  The exact solvability enables us to examine experimentally-accessible properties at both zero and finite $T$~\cite{PhysRevLett.112.207203,PhysRevLett.113.187201,PhysRevB.92.094439,PhysRevB.92.115127,PhysRevLett.113.197205,PhysRevB.92.115122,PhysRevLett.115.087203,yoshitake2016,Nasu2016nphys}, providing a good starting point to understand QSLs and transitions into different phases.

  To compare against candidate materials, such as iridates~\cite{PhysRevLett.108.127203,PhysRevLett.109.266406} and $\alpha$-RuCl$_3$~\cite{PhysRevB.91.094422,PhysRevB.90.041112}, it is essential to include residual interactions to study the competition between the Kitaev QSL and magnetically ordered phases~\cite{PhysRevLett.105.027204,PhysRevLett.110.097204,PhysRevLett.113.107201,PhysRevB.84.100406,PhysRevB.84.180407,1367-2630-16-1-013056,PhysRevB.88.035107,PhysRevB.90.155126,PhysRevB.92.184411,PhysRevB.93.155143,banerjee2016proximate,PhysRevB.93.174425}.
  This interest has created intense research activities to identify a trace of the Kitaev QSL \emph{at finite temperature} in a system whose ground state is on the ordered side though close to the QSL~\cite{PhysRevB.93.174425,PhysRevB.92.100403,banerjee2016proximate,PhysRevLett.114.147201}.
  However, as exemplified by the quantum Hall states, a phase bordering on the QSL is not limited to magnetically ordered states but can be another liquid-like state. Nevertheless, such a magnetic analog of a ``liquid-liquid (LL)'' transition remains elusive in the context of the Kitaev model, not to mention its signature at finite temperatures.

  In this Letter, we present for the first time an extension of the Kitaev model undergoing a topological LL transition at $T = 0$, and discuss the finite-$T$ phase diagram from the standpoint of fractionalization of quantum spins.
  Specifically, we consider the honeycomb-lattice Kitaev model with bond asymmetry by adding the Ising interaction.
  We find that the competition between the Kitaev QSL and the magnetically ordered phases in this model gives rise to a new intermediate liquid-like state.
  We show that the new nonmagnetic state is well described by a spin-nematic wavefunction without topological order.
  The phase diagram is obtained by using complementary methods: numerical calculations for finite-size clusters, the mean-field (MF) approximation, and the analyses of effective models.
  We show that the topological LL (i.e., QSL-nematic) transition is discontinuous, and the first-order transition line persists at finite $T$, terminating at a critical point.
  The LL transition is driven by the fractionalized excitations, $Z_2$ gauge fluxes, and the critical point locates deep inside the low-$T$ peculiar paramagnet, dubbed ``fractionalized paramagnet,'' which is set apart from the conventional paramagnet by a crossover driven by the spin fractionalization~\cite{PhysRevB.92.115127}.

\begin{figure*}[t]
\begin{center}
\includegraphics[width=2\columnwidth,clip]{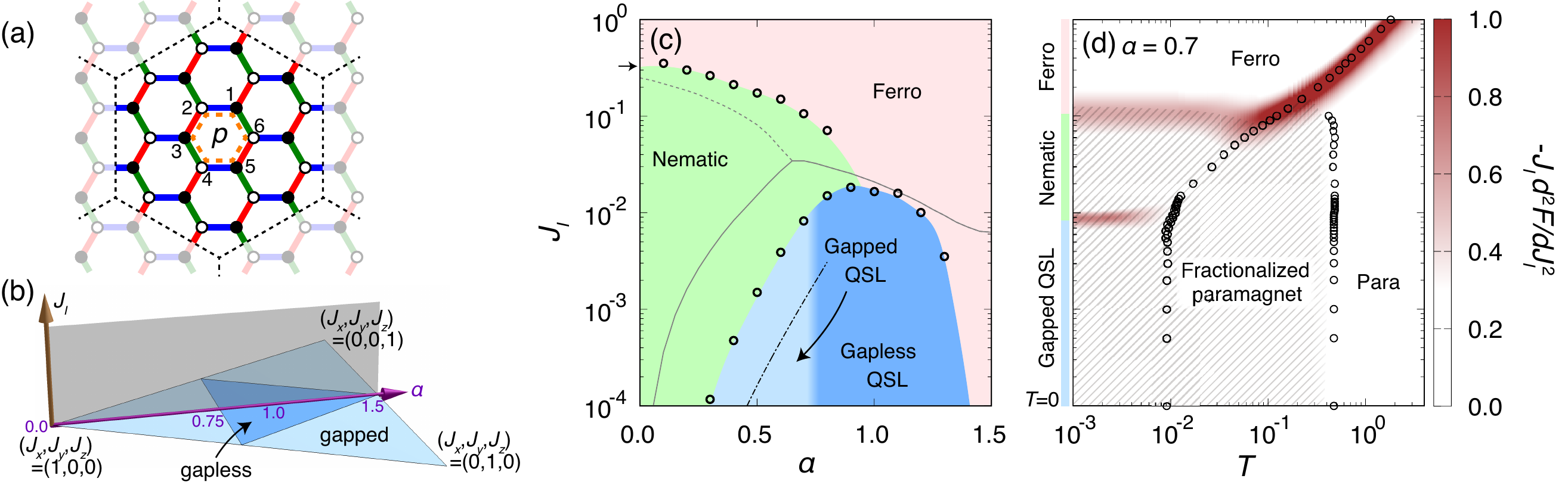}
\caption{
 (a) 24-site cluster of the honeycomb lattice with periodic boundary conditions, where blue, green, and red lines represent $x$-, $y$-, and $z$-bonds, respectively.
 The six-site loop with the dashed line represents $W_p = \sigma^z_1 \sigma^y_2 \sigma^x_3 \sigma^z_4 \sigma^y_5 \sigma^x_6$.
 (b) Ground-state phase diagram of the pure Kitaev model on the basal plane of $J_x+J_y+J_z=1$.
 The $\alpha$- and the vertical axes parametrize the bond asymmetry (see the text) and $J_I$, respectively.
  (c) Ground-state phase diagram of the Kitaev-Ising model obtained by the ED at $T=0$.
  The circles represent the phase boundary determined by the peaks of $-J_I d^2 E / d J_I^2$ (see Fig.~\ref{zeromoment}).
  The dashed-dotted line shows the asymptotic phase boundary, $J_I = J_y^2J_z^2/(16J_x^3)$, between the Kitaev and the spin-nematic liquids in the large-$J_x$ limit.
  The small arrow on the vertical axis indicates the Ising transition point $J_I = 0.3258(1)$ for $\alpha = 0$ (see the text).
 The solid (dotted) line shows the first- (second-)order phase boundary in the MF approximation.
  (d) Finite-$T$ phase diagram at $\alpha=0.7$ obtained by the TPQ state approach in the 24-site cluster in (a).
  The intensity map shows $-J_I d^2F/dJ_I^2$, the peaks of which imply the phase boundaries.
  The circles represent the peaks of the specific heat (see Fig.~\ref{Tdep}) and the hatched area corresponds to the fractionalized paramagnetic regime.
}
\label{lattice}
\end{center}
\end{figure*}

  We consider the Hamiltonian given by
\begin{eqnarray}
  {\cal H}=-\sum_{\gamma=x,y,z}\sum_{\means{jk}_\gamma}J_{\gamma} \sigma_j^\gamma\sigma_k^\gamma -J_I \sum_{\means{jk}} \sigma_j^z \sigma_k^z,
  \label{eq:H_KI}
\end{eqnarray}
where $\sigma_j^{x,y,z}$ denotes Pauli matrices representing the $S=1/2$ spin at site $j$, $J_\gamma$ is the Kitaev interaction on the $\gamma$-bond $\means{jk}_\gamma$ of the honeycomb lattice [see Fig.~\ref{lattice}(a)], and $J_I>0$ is the ferromagnetic Ising coupling for all the nearest neighbors (NNs); we call this the Kitaev-Ising model.
  We investigate not only the isotropic case $J_x=J_y=J_z$~\cite{PhysRevB.84.155121} but also the range of anisotropy covering both gapped and gapless QSL regimes.
Hereafter, we parametrize the Kitaev interactions as $J_x=1-2\alpha/3$ and $J_y=J_z=\alpha/3$ taking $J_x+J_y+J_z=1$ as the unit of energy ($J_\gamma \geq 0$).
  For $J_I = 0$, the gapless-gapped phase boundary is located at $\alpha=3/4$ [Fig.~\ref{lattice}(b)].

  We investigate the phase diagram of the model in Eq.~(\ref{eq:H_KI}) by several complementary methods.
  One is the numerical calculations of a small-size cluster: we perform the exact diagonalization (ED) at $T=0$ and the thermal pure quantum (TPQ) state approach for $T>0$~\cite{PhysRevLett.108.240401,PhysRevLett.111.010401,PhysRevB.90.121110} for the 24-site cluster shown in Fig.~\ref{lattice}(a).
  We also perform the MF calculation following Ref.~\cite{PhysRevB.84.155121}, which is based on a Majorana fermion representation of Eq.~(\ref{eq:H_KI})~\cite{PhysRevB.76.193101,PhysRevLett.98.087204,1751-8121-41-7-075001},
  \begin{eqnarray}
   \hspace{-4mm} {\cal H}=-i\sum_{\gamma=x,y}\sum_{\langle jk\rangle_\gamma} J_\gamma c_j c_k +J_z\sum_{\means{jk}_z}c_j c_k \bar{c}_j \bar{c}_k +J_I\sum_{\means{jk}}c_j c_k \bar{c}_j \bar{c}_k,\label{eq:1}
  \end{eqnarray}
  where $c$ and $\bar{c}$ represent Majorana operators, and $j$ and $k$ denote sites on different sublattices represented by filled and open circles, respectively, in Fig.~\ref{lattice}(a).
  For $J_I=0$, the Majorana fermions $\bar{c}$ are localized on $z$ bonds giving rise to %the static $Z_2$ gauge fields, because $\bar{c}_j \bar{c}_k$ on each $z$ bond commutes with ${\cal H}$.
the static $Z_2$ gauge fluxes, which commute with $\mathcal{H}$~\cite{Kitaev2006}; e.g., $W_p = \sigma_1^z\sigma_2^y\sigma_3^x\sigma_4^z\sigma_5^y\sigma_6^x = \bar{c}_2 \bar{c}_3 \bar{c}_5 \bar{c}_6$ [see Fig.~\ref{lattice}(a)].
  In addition, we analyze two effective models derived in appropriate limits, which provide complementary information on the thermodynamic limit.

\begin{figure}[t]
 \begin{center}
  \includegraphics[width=\columnwidth,clip]{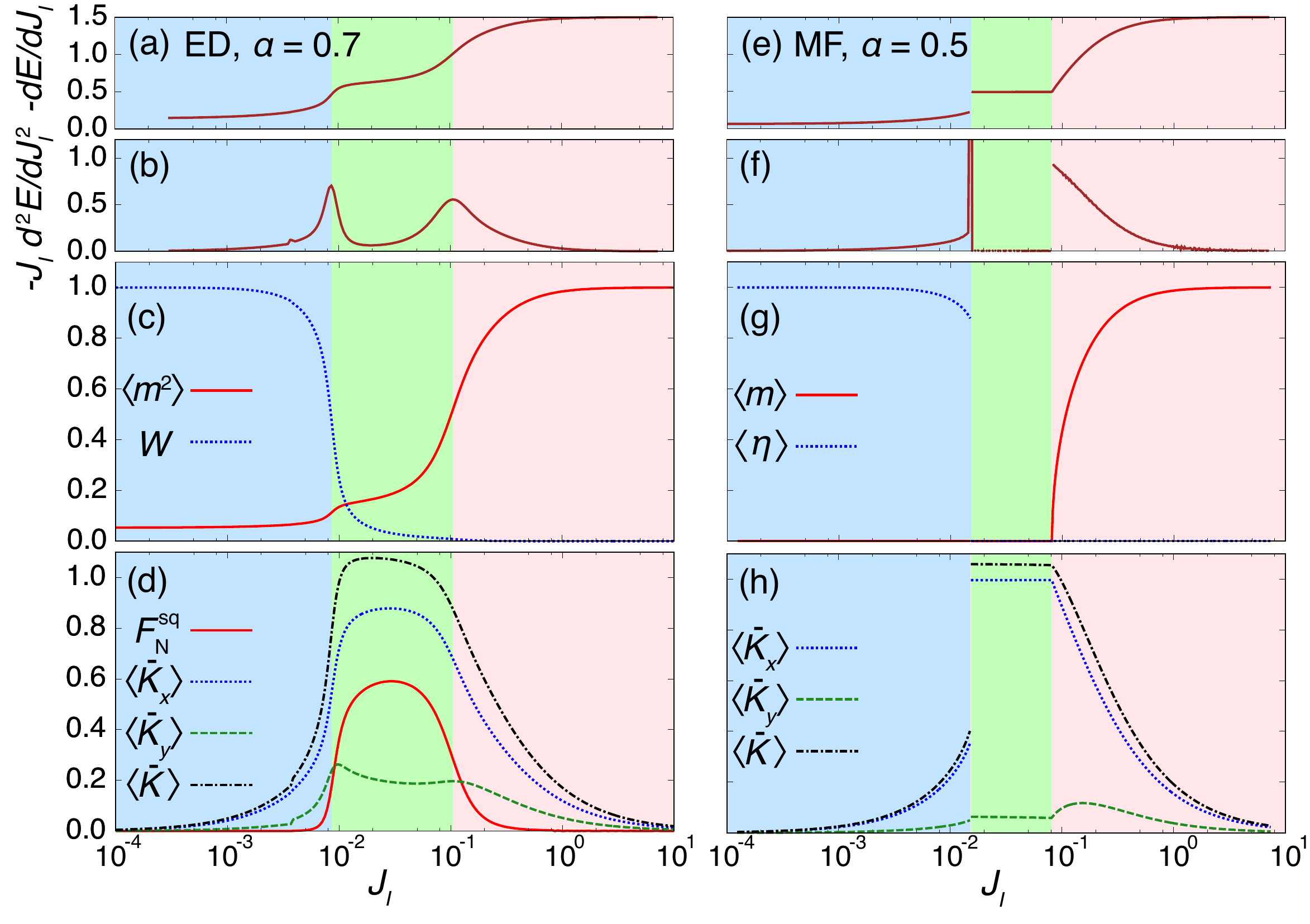}
  \caption{
  (a)--(d) Ground state properties at $\alpha = 0.7$ obtained with the ED in the 24-site cluster shown in Fig.~\ref{lattice}(a):
  (a) the first and (b) the second derivatives of the ground state energy in terms of $J_I$, (c) mean square of the magnetization $\means{m^2}$ and the flux density $W$, and (d) the dimensionless measures of kinetic energy of the Majorana fermions $\bar{c}$, $\means{\bar{K}_x}$, $\means{\bar{K}_y}$, $\means{\bar{K}} = \means{\bar{K}_x} + \means{\bar{K}_y}$, and the squared fidelity $\fidelity$ (see the text).
  (e)--(h) Majorana fermion MF results for the same or related observables (except for $\fidelity$) for $\alpha=0.5$: (e) $-dE/dJ_I$, (f) $-J_I d^2E/dJ_I^2$, (g)
    $\means{m}=(1/N)\sum_i\means{\sigma_i^z}$ and $\means{\eta}=(2/N)\sum_{\means{ij}_z}\means{\bar{c}_j\bar{c}_k}$, and (h) $\means{\bar{K}_x}$, $\means{\bar{K}_y}$, and $\means{\bar{K}}$. We note $W \approx \means{\eta}^2$ in the MF approximation because of the relation $W_p=\eta_r\eta_{r'}$ for each plaquette $p$~\cite{PhysRevB.76.193101,PhysRevLett.98.087204,1751-8121-41-7-075001}, where $r,r'$ represent two $z$ bonds of the plaquette $p$.
}
  \label{zeromoment}
 \end{center}
\end{figure}

  Figure~\ref{lattice}(c) shows the ground-state phase diagram obtained by the ED method on the $(\alpha,J_I)$ plane corresponding to the shaded plane in Fig.~\ref{lattice}(b)~\footnote{The gapped-gapless boundary for $J_I > 0$ seems almost insensitive to $J_I$ according to our MF calculation, while the ED for 24 spins is not efficient to clarify this subtlety.}.
  The QSL remains stable up to nonzero $J_I$, and a ferromagnetic phase appears for large enough $J_I$ as expected.
  Remarkably, however, we find another nonmagnetic phase between the QSL and the ferromagnetic phase.
  Figure~\ref{zeromoment}(a) shows the $J_I$-derivative of the ground state energy, $-dE/dJ_I$, evaluated with the ED calculation at $\alpha=0.7$.
  It shows steeper slopes around $J_{Ic1}\sim 0.01$ and $J_{Ic2}\sim 0.1$, which can be more clearly seen as peaks in $-J_Id^2 E/dJ_I^2$ [Fig.~\ref{zeromoment}(b)].
  Such peaks are known to be a good indicator of a phase transition even in small-size cluster calculations~\cite{PhysRevLett.105.027204}, suggesting three phases separated by transitions at $J_{Ic1}$ and $J_{Ic2}$.

  To characterize these phases, first we calculate the mean square of magnetization $\means{m^2}=\means{[(1/N)\sum_i\sigma_i^z]^2}$ and $W=|\means{W_p}|$. %, where $W_p=\sigma_1^z\sigma_2^y\sigma_3^x\sigma_4^z\sigma_5^y\sigma_6^x$ is the $Z_2$ flux on the plaquette $p$ with the site indices shown in Fig.~\ref{lattice}(a)~\cite{Kitaev2006}.
  For $J_I=0$, the exact ground state corresponds to $W=1$, while $\means{m^2} \to 1$ is expected as $J_I \to \infty$ because of the ferromagnetic order.
  The $J_I$ dependences of $\means{m^2}$ and $W$ for $\alpha = 0.7$ are presented in Fig.~\ref{zeromoment}(c).
  $W$ abruptly decreases from 1 to 0 around $J_{Ic1}$ with increasing $J_I$, which implies that the topological order is lost at $J_I = J_{Ic1}$, whereas the rapid increase of $\means{m^2}$ around $J_{Ic2}$ implies the onset of the ferromagnetic long-range order.
  The Majorana fermion MF calculations yield qualitatively similar results [Figs.~\ref{zeromoment}(e)--\ref{zeromoment}(g)], and the topology of the MF phase diagram is the same as the ED results [Fig.~\ref{lattice}(c)].
  Hence, neither $W$ nor $\means{m^2}$ can fully characterize the intermediate phase in the range $J_{Ic1}<J<J_{Ic2}$.

  To clarify the nature of the intermediate phase, a key observation is that the new phase is widely extended in the small  $\alpha$ regime ($J_x \gg J_\perp \equiv J_y = J_z$), as shown in Fig.~\ref{lattice}(c).
  This suggests that one can derive an effective model in the anisotropic limit.
  As Kitaev demonstrated~\cite{Kitaev2006}, such a large-$J_x$ effective model for $J_I = 0$ is the toric code appearing at the fourth order in $J_\perp$, which acts on the low-energy subspace spanned by direct products of $\ketbasic{\ra_j\ra_k}$ and $\ketbasic{\la_j\la_k}$ defined on each $x$-dimer $\means{jk}_x$, where $\ketbasic{\ra}$ and $\ketbasic{\la}$ are the eigenstates of $\sigma^x$ with eigenvalues $+1$ and $-1$, respectively.
  For small nonzero $J_I$, the leading $\mathcal{O}(J_I)$ contribution is to mutually flip $\ketbasic{\ra_j\ra_k}$ and $\ketbasic{\la_j\la_k}$ on each $x$ dimer due to the operator $\sigma_j^z \sigma_k^z$.
  Thus, to $\mathcal{O}(J_\perp^4/J_x^3,J_I)$, the effective Hamiltonian is the toric code in an effective transverse field.
  This perturbative argument suggests that the instability of the gapped QSL for large $J_x$ leads to a different ground state well approximated by
  \begin{eqnarray}
    \kets{\nematic} =\prod_{\means{jk}_x}
    \frac{1}{\sqrt{2}} \left( \ketbasic{\ra_j\ra_k} + \ketbasic{\la_j\la_k} \right),
    \label{eq:psiN}
  \end{eqnarray}
which has no magnetic moment but has a spin quadrupole moment constructed from the tensor operator $\sigma_j^\alpha \sigma_k^\beta$ ($\alpha,\beta=x,y,z$) for each $x$-bond $\means{jk}_x$, and hence is called the spin-nematic state~\cite{Lacroix2011C13}.
  Here, $\bras{\nematic}\sigma_j^y\sigma_k^y\kets{\nematic}=-1$ on the $x$ bonds $\means{jk}_x$ despite the absence of the interaction of $\sigma_j^y\sigma_k^y$ on the corresponding $x$ bonds in the Hamiltonian in Eq.~(\ref{eq:H_KI}).
  The quadrupole moment is not a spontaneous one but dictated by the symmetry of the bond-asymmetric Hamiltonian: $\kets{\nematic}$ is not a quadrupolar ordered state but just a quantum paramagnet.

  The spin-nematic nature is indeed confirmed by computing the squared fidelity $\fidelity = |\braketbasic{\nematic}{\psi_\text{gs}}|^2$, where $\ketbasic{\psi_\text{gs}}$ is the ground-state wave function.
  As shown in Fig.~\ref{zeromoment}(d), $\fidelity$ obtained by ED shows a substantial nonzero value only for $J_{Ic1} < J_I < J_{Ic2}$, which indicates that the intermediate phase is well described by $\kets{\nematic}$.
  We note that $\fidelity$ is considerably reduced from the unity due to quantum corrections induced by the next-to-leading $\mathcal{O}(J_\perp J_I / J_x)$ and higher order effects~\cite{suppl}.

  In the spin-nematic quantum paramagnet, the quadrupole moment enhances
  $
  \means{\bar{K}_x} = -
  (2/N)
  \sum_{\means{jk}_x}\means{\sigma_j^y\sigma_k^y}
  $.
  Interestingly, $\means{\bar{K}_y} = -(2/N)\sum_{\means{jk}_y}\means{\sigma_j^x\sigma_k^x}$ is also enhanced due to quantum fluctuations, as shown in Fig.~\ref{zeromoment}(d).
  These are the dimensionless measures of the kinetic energy of the Majorana fermions $\bar{c}$ producing the static $Z_2$ gauge fluxes for $J_I = 0$, because $\bar{K}_{x(y)} = (2/N) \sum_{\means{jk}_{x(y)}}i\bar{c}_j\bar{c}_k$.
  Such enhanced kinetic energy of $\bar{c}$ is also captured by the Majorana fermion MF calculation [Fig.~\ref{zeromoment}(h)], which is due to the term, $-J_I\means{c_j c_k} \bar{c}_j \bar{c}_k$.
  Thus, $\bar{c}$ is delocalized in the spin-nematic region, while it is completely localized in the Kitaev limit.

  As both the Kitaev QSL and the spin-nematic states break no symmetry, the conventional Landau theory dictates no phase transitions between them.
  Nevertheless, the two states can be distinguished in terms of topology; thus we expect a topological phase transition at zero $T$.
  From the mapping to the transverse-field toric code, which undergoes a first-order topological transition~\cite{PhysRevB.80.081104}, we conclude that the QSL-nematic LL transition is also of first order, at least, in the large $J_x$ limit.
  Indeed, our MF calculation indicates the first-order nature [see Figs.~\ref{zeromoment}(e) and \ref{zeromoment}(f)], and our ED results also show a rather sharp change in the energy at the phase boundary [Figs.~\ref{zeromoment}(a) and \ref{zeromoment}(b)].
  In addition, the asymptotic phase boundary in the large-$J_x$ limit can be predicted as $J_I\sim J_y^2J_z^2/(16J_x^3)$ based on the same mapping~\cite{PhysRevB.80.081104}, which is plotted as the dashed-dotted line in Fig.~\ref{lattice}(c).
  The deviation of our ED results from this expected behavior is semi-quantitatively accounted by considering the finite-size effect~\cite{suppl}, similar to the one in the pure Kitaev model~\cite{kells2009}.
  Thus, the large-$J_x$ analysis provides crucial information on the nature of the topological LL transition in the thermodynamic limit, complementary to the ED and MF results.

  Regarding the nematic-ferromagnetic transition, both the ED [Figs.~\ref{zeromoment}(a) and \ref{zeromoment}(b)] and the MF approximation [Figs.~\ref{zeromoment}(e) and \ref{zeromoment}(f)] suggest a continuous transition.
  In this case also, complementary information on the thermodynamic limit is brought by the analysis of an effective model for $J_y = 0$ and $J_I>0$~\cite{suppl}.
  The analysis predicts a continuous transition at $J_I=0.3285(1)$ for $\alpha=0$ ($J_y=J_z=0$), consistent with the ED results [see Fig.~\ref{lattice}(c)].

  Given the first-order topological transition at $T=0$, an interesting question must concern the fate of the LL transition at finite $T$.
  We determine the finite-$T$ phase diagram [Fig.~\ref{lattice}(d)] by calculating $-J_I d^2F/dJ_I^2$ ($F$ is the free energy), which is an extension of the approach for the $T=0$ case.
  The $T$ dependence is evaluated by the TPQ state approach in the 24-site cluster.
  While the ferromagnetic phase is surrounded by the peak of $-J_I d^2F/dJ_I^2$ associated with the symmetry breaking, the topological LL transition at $T = 0$ yields a first-order transition line persisting at $T>0$, as suggested by the peak emanating from it.
  This line disappears at finite $T\sim 10^{-2}$, indicating a critical point.
  The persistence of the first-order transition line at $T>0$ is supported by the effective model for large $J_x$, which retains the same thermodynamic feature
~\footnote{More precisely, the transverse-field toric code can be further mapped via the Xu-Moore model for the $p + ip$ superconductor arrays~\cite{PhysRevLett.93.047003} to a square-lattice compass model~\cite{PhysRevB.71.195120}, in which the first-order transition line terminating at a finite-$T$ critical point was confirmed by using quantum Monte Carlo simulations ~\cite{PhysRevLett.98.256402,PhysRevB.78.064402}.}.

\begin{figure}[t]
\begin{center}
\includegraphics[width=\columnwidth,clip]{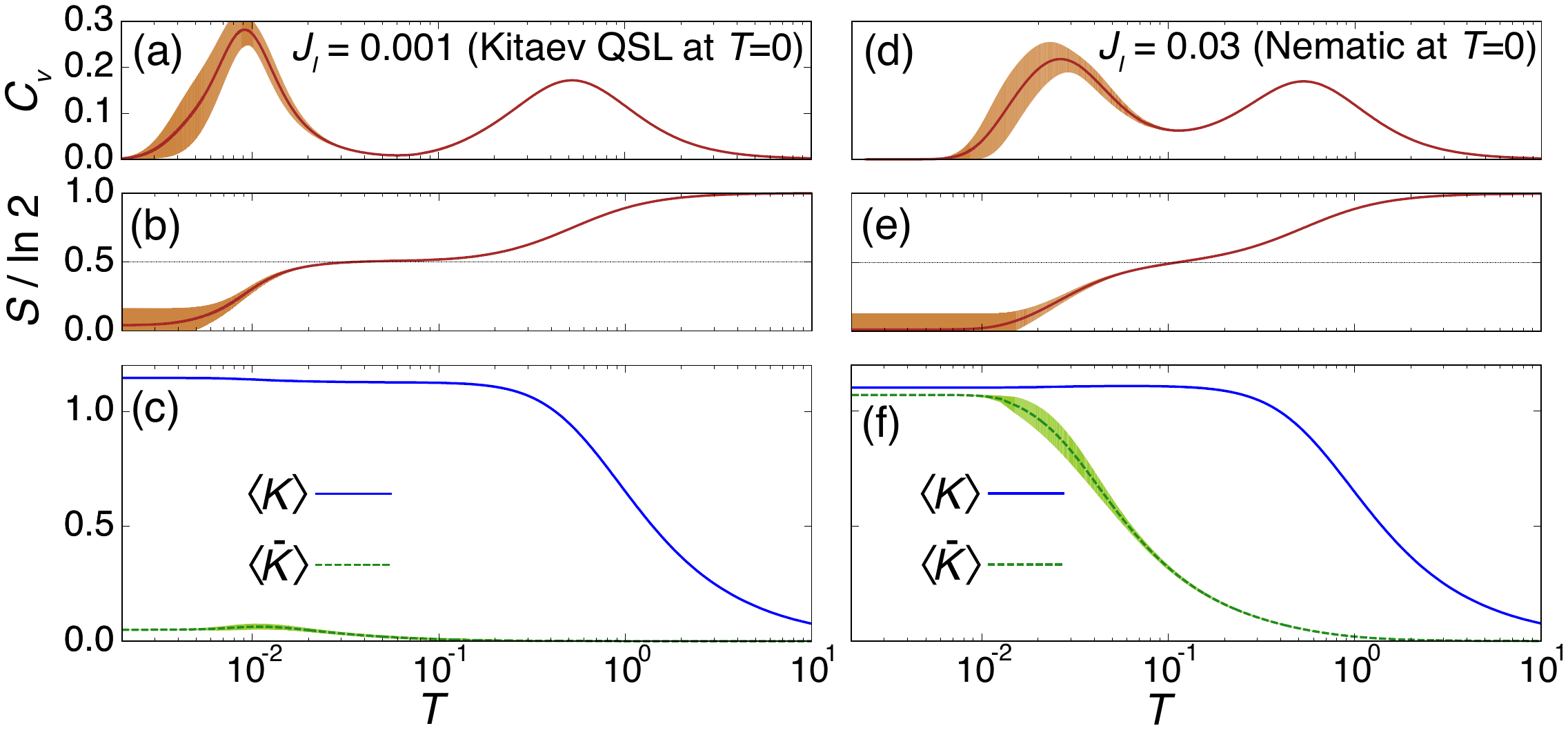}
\caption{
  (a)--(c) $T$ dependences of (a) the specific heat, (b) normalized entropy, and (c) the dimensionless measures of kinetic energy $\means{K}$ and $\means{\bar{K}}$ of $c$ and $\bar{c}$, respectively, at $J_I=0.001$ and $\alpha=0.7$, where the ground state is the Kitaev QSL; $K=(2/N)\sum_{\gamma=x,y}\sum_{\means{jk}_\gamma}\sigma_j^\gamma\sigma_k^\gamma = (2/N)\sum_{\gamma=x,y}\sum_{\means{jk}_{\gamma}}i c_j c_k$.
  (d)--(f) Corresponding data for $J_I=0.03$ and $\alpha=0.7$, where the ground state is spin-nematic.
  The shaded broadening represents errors associated with the TPQ state approach.
}
\label{Tdep}
\end{center}
\end{figure}

  Notably, the critical point lies well inside the ``fractionalized paramagnet'' inherited from the pure Kitaev model.
  In the pure Kitaev model ($J_I=0$), the fractionalization is known to affect  the thermodynamics: the Majorana fermions $c$ release their entropy a half of $\ln 2$ per spin at a high $T\sim 0.5$, while the $Z_2$ fluxes release the remaining half at very low $T\lesssim 10^{-2}$~\cite{PhysRevB.92.115122}.
  The fractionalized paramagnet originally refers to the low-$T$ regime set apart from the conventional paramagnet by the high-$T$ crossover.
  In the present study, we find that the peculiar regime remains intact up to a finite deviation from the pure Kitaev model.
  As shown in Fig.~\ref{lattice}(d), the crossover line determined by the higher-$T$ broad peak in the specific heat associated with the entropy release of $(1/2)\ln 2$ continue for $J_I > 0$ [see Figs.~\ref{Tdep}(a), \ref{Tdep}(b), \ref{Tdep}(d), and \ref{Tdep}(e)], reaching the region where the ground state is the spin-nematic state.
  Also, as in the Kitaev limit~\cite{PhysRevLett.113.197205,PhysRevB.92.115122}, the kinetic energy of the Majorana fermion $c$ is enhanced below the higher-$T$ crossover at $T\sim 0.5$ [see Figs.~\ref{Tdep}(c) and \ref{Tdep}(f)].
  Thus, the LL critical point at $T\sim 10^{-2}$ locates deep inside the fractionalized paramagnetic regime, and the two liquid regions are adiabatically connected at finite $T$ through the fractionalized paramagnetic regime by bypassing the critical point.
  This does not contradict with the Landau theory as both liquids preserve the full global symmetry of ${\cal H}$.

  The low-$T$ crossover corresponds to the emergence of the distinct characterizations of the two liquids, and the appearance of the LL critical point at this $T$ scale is its direct consequence.
  Here, the Majorana fermions $\bar{c}$ take the leading role.
  To see this, we show the dimensionless measure of their kinetic energy $\bar{K} = \bar{K}_x + \bar{K}_y$ with $\bar{K}_{x,y}$ defined previously.
  When the ground state is the Kitaev QSL, $\means{\bar{K}}$ is very small at any $T$ [Fig.~\ref{Tdep}(c)].
  On the other hand, when the ground state is spin-nematic, $\means{\bar{K}}$ starts to develop around the low-$T$ crossover [Fig.~\ref{Tdep}(f)].
  These observations are consistent with the fact that the two spin liquids are of different origins, as discussed above for $T=0$: while the Majorana fermions $\bar{c}$ are localized at any $T$ in the Kitaev QSL, they become delocalized in the spin-nematic liquid below the low-$T$ crossover.
Note that the lower crossover temperature increases with $J_I$ almost linearly in the nematic case [Fig.~\ref{lattice}(d)]. This is consistent with the MF picture that the itinerant behavior of $\bar{c}$ is driven by $-J_I \means{c_j c_k} \bar{c}_j \bar{c}_k$.
  Thus, separated by the first-order transition line, $\bar{c}$ is localized (delocalized) for smaller (larger) $J_I$, indicating that the driving force of the LL transition is delocalization of emergent fractional particles $\bar{c}$.

  In summary, we investigated the Kitaev-Ising model at both zero and finite $T$ and found the liquid-liquid transition between the Kitaev QSL and the newly-identified spin-nematic quantum paramagnet.
  The QSL-nematic transition is of first order, driven by delocalization of $\bar{c}$ composing $Z_2$ gauge fluxes.
  While the two liquids can be distinguished at $T = 0$ in terms of topological order, we demonstrated that the first-order transition line persists at finite $T$ and terminates inside the fractionalized paramagnetic regime.
  We emphasize that the Kitaev-Ising Hamiltonian is not a prerequisite for the proposed liquid-liquid transition; similar transitions may occur in other extended Kitaev models, such as the Kitaev-Heisenberg model~\cite{PhysRevLett.105.027204} and the $J$-$K$-$\Gamma$ model~\cite{PhysRevLett.112.077204}, with large bond asymmetry.
  We hope our work will promote further studies for unveiling new quantum phases proximate to the Kitaev QSL and stimulate experimental efforts on candidate materials.

\begin{acknowledgments}
This work is supported by Grants-in-Aid for Scientific Research under Grants No. JP24340076, JP26800199, JP15K13533, JP16K17747, and JP16H02206.
Parts of the numerical calculations were performed in the supercomputing systems in ISSP, the University of Tokyo.
The TPQ state calculations were performed using the ${\cal H}\Phi$ package~\cite{HPhipre}.
\end{acknowledgments}

% % % % % % % % % % % % % % % % % % % % % % % % % % % % % % % % % % %
%                                                                   %
% Supplmental Material                                              %
%                                                                   %
% Please leave this as it is to generate a file for arXiv and       %
% comment out the following when submitting the manuscript to APS   %
%                                                                   %
% % % % % % % % % % % % % % % % % % % % % % % % % % % % % % % % % % %
%                                                                   %
\appendix                                                           %
\vspace{15pt}                                                       %
\begin{center}                                                      %
{\bf ---Supplemental Material---}                                  %
\end{center}                                                        %
\input{SI_v3.0_body.tex}                               %
%                                                                   %
% % % % % % % % % % % % % % % % % % % % % % % % % % % % % % % % % % %

\bibliography{refs}
% \nocite{apsrev41Control}
% \bibliographystyle{my-apsrev4-1}
% \bibliography{my-refcontrol,refs}

\end{document}

%% file: SI_v3.0_body.tex
\setcounter{figure}{0}
\setcounter{equation}{0}
\setcounter{table}{0}
\renewcommand{\thefigure}{S\arabic{figure}}
\renewcommand{\theequation}{S\arabic{equation}}
\renewcommand{\thetable}{S\Roman{table}}

\section{I.~~ Finite-size effect near the QSL-nematic transition}

  In this section, we explain the origin of the rather strong finite-size effect seen in the QSL-nematic transition in Fig.~1(c) in the main text.
  While this transition can be best understood through the mapping to the transverse-field toric code model as discussed in the main text, additional topological boundary terms appear in finite-size clusters from the perturbation processes passing through the boundary of the cluster.
  The additional terms are also conserved quantities at $J_I = 0$ as the usual toric code terms, hence lowering the ground state energy in the QSL phase.
  Consequently, it is expected that the QSL region can be overestimated in finite-size clusters.
  Below we confirm this expectation, showing explicitly that the strong finite-size effect exists in the $N = 24$ cluster [Fig.~1(a)] used to construct our phase diagram in the main text, due to such boundary terms appearing at the same order as the usual toric code terms.

  We assume $J_{x} \gg J_{\perp} \equiv J_{y} = J_{z} \gg J_I >0$.
  In the thermodynamic limit, the effective Hamiltonian to $\mathcal{O}(J_{\perp}^4/J_x^3, J_{\perp} J_I/J_x)$ comprises the four-pseudospin toric code term $J_\text{TC}$, the effective transverse field term $h_\text{eff}$, and the Ising term $J^\tau_z$ on the $y$ bonds, as
\begin{align}
  \mathcal{H}_\text{\,eff,\,bulk}
        =& -J_\text{TC} \sum_{R}
           \tau^y_R \tau^z_{R+\mathbf{a}} \tau^y_{R+\mathbf{a}+\mathbf{b}} \tau^z_{R+\mathbf{b}}
  - h_\text{eff} \sum_{R} \tau_R^x
  \notag\\
  &+ J^\tau_{z} \sum_{R} \tau_R^z \tau_{R+\mathbf{b}}^z + c(J_x, J_{\perp}, J_I),
  \label{eq:toric}
\end{align}
with $J_\text{TC} = J_{\perp}^4 / (16 J_x^3)$~\cite{Kitaev2006}, $h_\text{eff} = J_I$, and $J^\tau_{z} = J_{\perp} J_I / (2J_x)$, where $R$ runs over the $J_x$ dimers. $\tau^{x,y,z}$ are the pseudospin Pauli matrices acting on the low-energy subspace spanned by direct products of $\ket{\ra\ra}_R$ and $\ket{\la\la}_R$, namely,
\begin{align}
        \tau^z = \begin{pmatrix} 1 & 0 \\ 0 & -1 \end{pmatrix},~~
        \tau^x = \begin{pmatrix} 0 & 1 \\ 1 &  0 \end{pmatrix},~~
        \tau^y = \begin{pmatrix} 0 & -i \\ i & 0 \end{pmatrix},
\end{align}
in this dimer basis.
  Here, $\kets{\ra\ra}_R=\kets{\ra_j\ra_k}$ for the dimer $R$ composed of sites $j$ and $k$ in the main text.
  For $J_I = 0$, the ground state energy per unit cell (of the honeycomb lattice) is $E_\text{\,eff,\,bulk}^{J_I = 0} = -J_{\text{TC}} + \mathcal{O}(J_{\perp}^6 /J_x^5)$ apart from the constant term in Eq.~\eqref{eq:toric},
\begin{align}
  c(J_x, J_{\perp}, J_I) =& -J_x - J_{\perp}^2 / (4 J_x) - (J_{\perp}^2 + 2 J_\perp J_I) / (4 J_x)
  \notag\\
  &+ J_{\perp}^{4} / (32 J_x^3) + \mathcal{O}(J_{\perp}^6/J_x^5,(J_{\perp}/J_x)^2J_I).
\end{align}
  For $J_z^\tau = 0$, the Hamiltonian \eqref{eq:toric} coincides with the transverse-field toric code studied in Ref.~\cite{PhysRevB.80.081104}.

\begin{figure}[t]
  \begin{center}
    \includegraphics[width=0.65\hsize,bb=0 0 283 544]{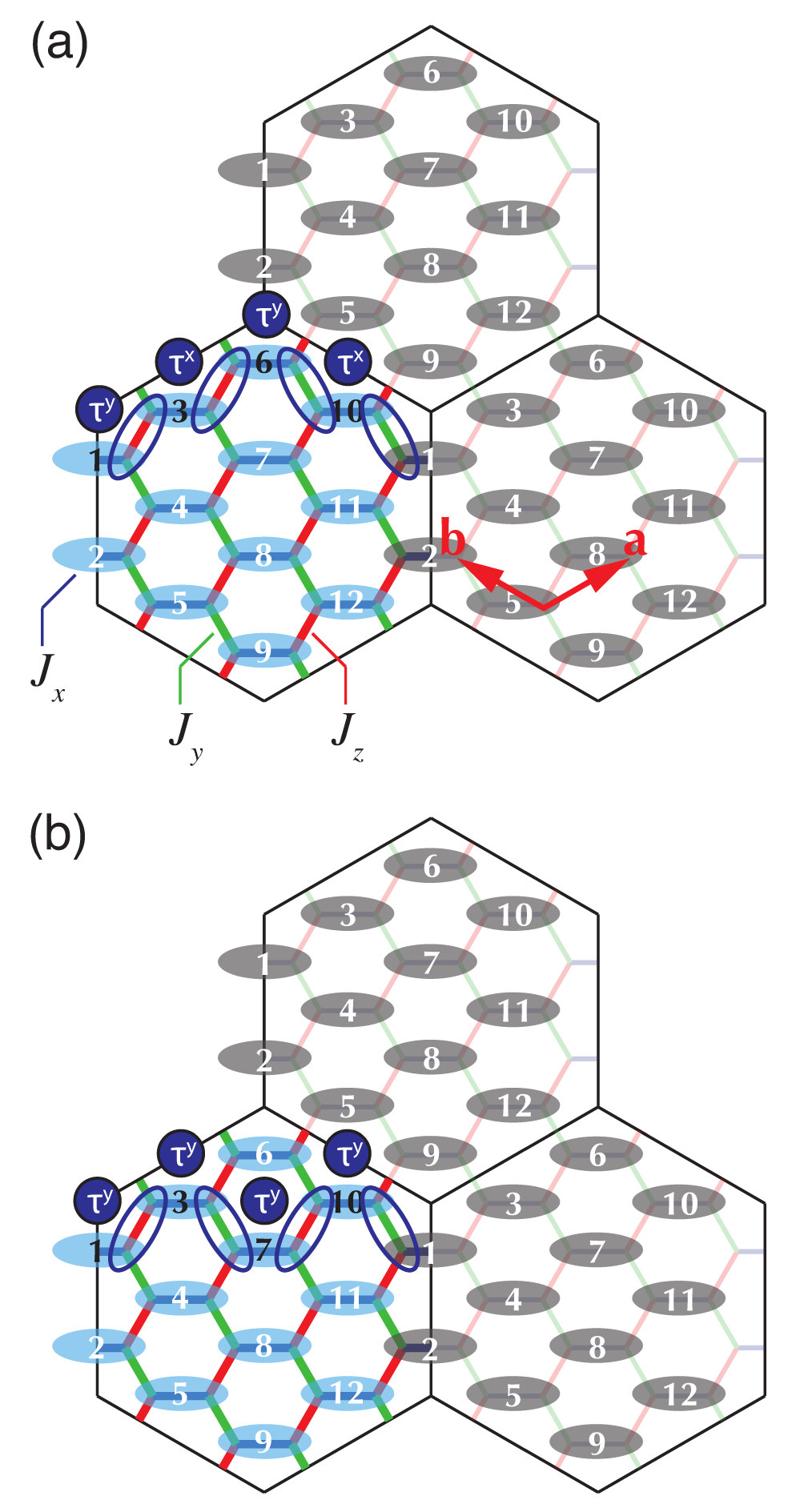}
  \end{center}
  \caption{
    \label{fig:paths}
    Schematic pictures of the additional perturbation processes in
    the $N = 24$ cluster.
    (a) and (b) correspond to Eqs.~(\ref{eq:FSE1}) and (\ref{eq:FSE2}), respectively.
    See Tables~\ref{table:FSE1} and \ref{table:FSE2} for the complete lists of the dimers
    belonging to the categories (a) and (b), respectively.
  }
\end{figure}

\begin{table}
\begin{center}
  \caption{
    The dimer-site indices $R_1$--$R_4$ contributing to $\mathcal{H}_\text{\,eff,\,boundary\,(1)}^{R_1 R_2 R_3 R_4}$ in Eq.~\eqref{eq:FSE1}.
  }
  \vspace{4pt}
  \scalebox{1}{
    \begin{tabular}{cccc}
    \toprule
    \parbox[t]{45pt}{$R_1$} & \parbox[t]{45pt}{$R_2$} & \parbox[t]{45pt}{$R_3$} & \parbox[t]{45pt}{$R_4$} \\
    \midrule
    1  &  3  &  6  & 10 \\
    2  &  4  &  7  & 11 \\
    6  &  5  &  8  & 12 \\
    7  & 10  &  9  &  3 \\
    8  & 11  &  1  &  4 \\
    9  & 12  &  2  &  5 \\
    3  &  6  &  5  &  9 \\
    4  &  7  & 10  &  1 \\
    5  &  8  & 11  &  2 \\
    10 &  9  & 12  &  6 \\
    11 &  1  &  3  &  7 \\
    12 &  2  &  4  &  8 \\
    \bottomrule
  \end{tabular}
  }
  \label{table:FSE1}
\end{center}
\end{table}

\begin{table}
\begin{center}
  \caption{
    The dimer-site indices $R_1$--$R_4$ contributing to $\mathcal{H}_\text{\,eff,\,boundary\,(2)}^{R_1 R_2 R_3 R_4}$ in Eq.~\eqref{eq:FSE2}.
  }
  \vspace{4pt}
  \scalebox{1}{
    \begin{tabular}{cccc}
    \toprule
    \parbox[t]{45pt}{$R_1$} & \parbox[t]{45pt}{$R_2$} & \parbox[t]{45pt}{$R_3$} & \parbox[t]{45pt}{$R_4$} \\
    \midrule
    1  &  3  &  7  & 10 \\
    2  &  4  &  8  & 11 \\
    6  &  5  &  9  & 12 \\
    3  &  6  & 10  &  9 \\
    4  &  7  & 11  &  1 \\
    5  &  8  & 12  &  2 \\
    \bottomrule
  \end{tabular}
  }
  \label{table:FSE2}
\end{center}
\end{table}

\begin{figure*}[t]
  \begin{center}
    \includegraphics[width=\hsize, bb=0 0 1364 371]{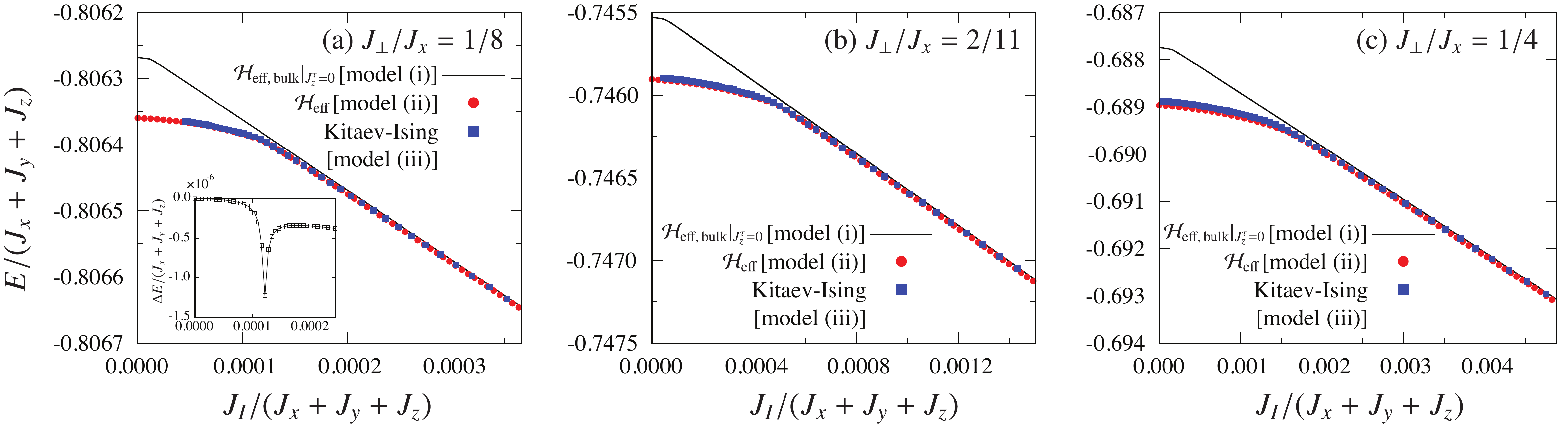}
  \end{center}
  \caption{
    \label{fig:comparison}
 Ground-state energy density $E$ for (a) $J_{\perp} / J_x = 1/8$, (b) $J_{\perp} / J_x = 2/11$, and (c) $J_{\perp} / J_x = 1/4$ ($\alpha = 0.3$, $0.4$, and $0.5$, respectively) evaluated with the different approaches (i)--(iii) (see the text).
 The inset in (a) shows the change $\Delta E = E\hspace{1pt}\rvert_{J^\tau_z \ne 0} - E\hspace{1pt}\rvert_{J^\tau_z = 0}$ induced by the $J^\tau_z$ term in the extended toric code Hamiltonian in a transverse magnetic field [Eq.~\eqref{eq:toric}].}
\end{figure*}

\begin{figure*}[t]
  \begin{center}
    \includegraphics[width=\hsize, bb=38 0 1343 370]{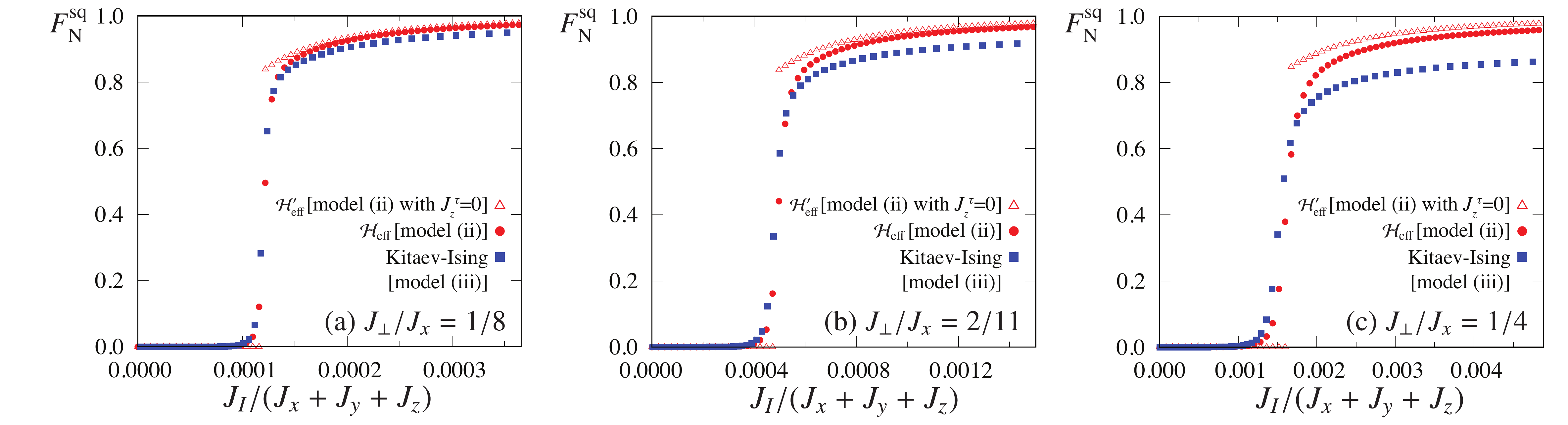}
  \end{center}
  \caption{
    \label{fig:comparison:Fsq}
    Squared fidelity $\fidelity$ of the ground state with respect to the direct product of the bond nematic states (see the text and Eq.~(3) of the main text). We show the comparisons among two variants of the approach (ii), namely, $\mathcal{H}_\text{\,eff}$ and $\mathcal{H}'_\text{\,eff}$ (see the text), and the Kitaev-Ising model for (a) $J_{\perp} / J_x = 1/8$, (b) $J_{\perp} / J_x = 2/11$, and (c) $J_{\perp} / J_x = 1/4$ ($\alpha = 0.3$, $0.4$, and $0.5$, respectively).
  }
\end{figure*}

In the $N = 24$ cluster shown in Fig.~1(a) or Fig.~\ref{fig:paths}, there are additional
contributions at the same order as the toric code terms arising from
the perturbation processes through the boundary of the cluster, as illustrated in Figs.~\ref{fig:paths}(a) and \ref{fig:paths}(b).
The perturbation corresponding to Fig.~\ref{fig:paths}(a) gives rise to the additional term,
\begin{align}
  \mathcal{H}_\text{\,eff,\,boundary\,(1)}^{R_1 R_2 R_3 R_4} = 5 J_\text{TC} \tau^y_{R_1} \tau^x_{R_2} \tau^y_{R_3} \tau^x_{R_4},
  \label{eq:FSE1}
\end{align}
where $R_1$--$R_4$ are the dimer sites contributing to the perturbation process; we summarize the list of $R_1$--$R_4$ in Table~\ref{table:FSE1}.
Similarly, the perturbation corresponding to Fig.~\ref{fig:paths}(b) leads to
\begin{align}
  \mathcal{H}_\text{\,eff,\,boundary\,(2)}^{R_1 R_2 R_3 R_4} = -5 J_\text{TC} \tau^y_{R_1} \tau^y_{R_2} \tau^y_{R_3} \tau^y_{R_4},
  \label{eq:FSE2}
\end{align}
with the associated sites $R_1$--$R_4$ in this category are summarized in Table~\ref{table:FSE2}.
All of these additional terms (18 terms in total) are conserved binary quantities, which we call the topological boundary terms.
Thus, the ground-state energy density in this cluster at $J_I = 0$ is lowered by
\begin{align}
\Delta E_\text{\,eff,\,bulk}^{J_I = 0}
= -\frac{15}{2}J_{\text{TC}} + \mathcal{O}(J_{\perp}^6 /J_x^5),
\end{align}
which in fact dominates the bulk contribution from Eq.~(\ref{eq:toric}).

  In Fig.~\ref{fig:comparison}, we compare the ground state energy per unit cell for several values of $J_{\perp} / J_x$ as a function of $J_I$ evaluated with different approaches:
  (i) the 10th-order perturbation theory presented in Ref.~\cite{PhysRevB.80.081104} for the transverse-field toric code model $\mathcal{H}_\text{\,eff,\,bulk}\rvert_{J^\tau_z = 0}$, which is expected to be accurate enough (the $J^\tau_z$ term gives a negligible contribution as discussed later),
  (ii) the effective Hamiltonian including the topological boundary terms $\mathcal{H}_\text{\,eff} = \mathcal{H}_\text{\,eff,\,bulk} + \sum_{(R_1,R_2,R_3,R_4)}\mathcal{H}_\text{\,eff,\,boundary\,(1)}^{R_1 R_2 R_3 R_4} + \sum_{(R_1,R_2,R_3,R_4)}\mathcal{H}_\text{\,eff,\,boundary\,(2)}^{R_1 R_2 R_3 R_4}$ for $N / 2 = 12$ pseudospins derived from the $N = 24$ cluster, and (iii) the original Kitaev-Ising model on the $N = 24$ cluster.
  We find that for all the cases shown in Figs.~\ref{fig:comparison}(a)--\ref{fig:comparison}(c), the results of (ii) and (iii) agree very well with each other, demonstrating the validity of our strong coupling approach including the topological boundary terms.
  Meanwhile, these results largely deviate from the bulk effective model (i) especially in the Kitaev QSL region for small $J_I$, where their ground state energies are lowered because of the boundary effect.
  In the main text, the phase boundary between the QSL and the spin-nematic phases is estimated from the first peak in $-J_I d^2 E / dJ_I^2$ [see Fig.~2(b)], which corresponds to the kinks seen in $E$ as a function of $J_I$ in Figs.~\ref{fig:comparison}(a)--\ref{fig:comparison}(c).
  Because $E$ turns out to be almost insensitive to the boundary effect in the spin-nematic phase for large $J_I$, the lowering of the energy in (ii) and (iii) for small $J_I$ results in the shift of the critical point estimated in the present finite-size cluster to larger values of $J_I$, hence overestimating the QSL phase.

However, because the boundary terms like Eqs.~\eqref{eq:FSE1} and \eqref{eq:FSE2} become increasingly higher order corrections on larger clusters and thus are negligible in the thermodynamic limit, this result indicates that Eq.~\eqref{eq:toric} is a very good effective theory in the regime of the QSL-nematic transition.
Thus, the asymptotic form of the QSL-nematic phase boundary mentioned in the main text corresponds to not the kinks in (ii) or (iii) but rather to the one in (i) in Fig.~\ref{fig:comparison}.

  We note that the effect of the additional $J^\tau_{z}$ term is almost negligible in usual observables like energy.
  Indeed, the same degree of agreement can be achieved also by considering ${\mathcal{H}}'_\text{\,eff} \equiv \mathcal{H}_\text{\,eff}\bigr\rvert_{J^\tau_z = 0} = \mathcal{H}_\text{\,eff,\,bulk}\bigr\rvert_{J^\tau_z = 0} + \sum_{(R_1,R_2,R_3,R_4)}\mathcal{H}_\text{\,eff,\,boundary\,(1)}^{R_1 R_2 R_3 R_4} + \sum_{(R_1,R_2,R_3,R_4)}\mathcal{H}_\text{\,eff,\,boundary\,(2)}^{R_1 R_2 R_3 R_4}$, as illustrated in the inset of Fig.~\ref{fig:comparison}(a).

  On the other hand, the $J^\tau_{z}$ term and even higher order corrections are important to reproduce the behavior of the squared fidelity $\fidelity$ in the Kitaev-Ising model (see Fig.~\ref{fig:comparison:Fsq}).
  It is found that $\fidelity$ in the spin nematic phase of the Kitaev-Ising model is more reduced than that of the transverse-field toric code for $J^\tau_{z} = 0$ with the finite-size term ($\mathcal{H}_\text{eff}\rvert_{J^\tau_{z}=0}$) and the $J^\tau_{z}$ term modifies $\fidelity$ in the right direction in the effective model.
  Still, discrepancies are left, which are more significant for larger values of $\alpha$, implying that $\fidelity$ is sensitive to the accuracy of the effective model description.

\section{II.~~ Reduced model for the nematic-magnetic transition}
  To discuss the transition between the spin nematic and the ferromagnetic phases, we construct a reduced model assuming $J_y=0$ and $J_I>0$.
  In this case, $\sigma_j^z\sigma_k^z$ on each $x$ bond is conserved, and the reduced Hamiltonian is block-diagonalized in subspaces specified by the eigenvalues of $\sigma_j^z\sigma_k^z$ on each $x$ bond.
  By introducing a new $x$-bond basis $\ketbasic{\Up}_{R} = \ketbasic{\up_j \up_k}\: (\ketbasic{\up_j \dw_k})$ and $\ketbasic{\Dw}_{R} = \ketbasic{\dw_j \dw_k}\: (\ketbasic{\dw_j \up_k})$ for $\sigma_j^z\sigma_k^z = +1\: (-1)$ on each $x$-bond $R = \means{jk}_x$ (here $\ketbasic{\up}$ and $\ketbasic{\dw}$ are the eigenstates of $\sigma^z$), the Hamiltonian in a subspace $V$ characterized by a configuration of $\{\sigma_j^z\sigma_k^z=\pm 1\}$ for $x$ bonds is given by
\begin{align}
{\cal H}_{\rm eff}^V&=\tilde{\cal H}_{\rm eff}^V + \tilde{E}^V,
\end{align}
where
\begin{align}
\tilde{\cal H}_{\rm eff}^V=&-J_I\sum_{\means{RR'}_y} c_{RR'}^V\tilde{\tau}_R^x \tilde{\tau}_{R'}^x
- (J_z+J_I) \sum_{\means{RR'}_z} c_{RR'}^V\tilde{\tau}_R^x \tilde{\tau}_{R'}^x\notag\\
&- J_x \sum_{R} (\tilde{\tau}_R^z+1),
\\
\tilde{E}^V&=-J_I \left(\frac{N}{2}-2n\right)+\frac{J_xN}{2}.
  \label{eq:EffHami}
\end{align}
  Here, the pseudospin Pauli matrices $\tilde{\tau}_R^{x,y,z}$ are given for the bases $\ketbasic{\Ra}_R=\frac{1}{\sqrt{2}}(\ketbasic{\Up}_{R}+\ketbasic{\Dw}_{R})$ and $\ketbasic{\La}_R=\frac{1}{\sqrt{2}}(\ketbasic{\Up}_{R}-\ketbasic{\Dw}_{R})$, $\means{RR'}_y$ ($\means{RR'}_z$) stands for a NN $x$-dimer sites connected by a $y$ ($z$) bond, and $n$ is the number of $x$ bonds with $\sigma_j^z\sigma_k^z = -1$.
  The coefficient $c_{RR'}^V$ taking $\pm 1$ for each $y$ or $z$ bond is given by $c_{RR'}^V=\brabasic{\psi_{\rm up}^V}\sigma_j^z \sigma_k^z\ketbasic{\psi_{\rm up}^V}$, where $j$ and $k$ are NN sites on the corresponding $y$ or $z$ bond of the original honeycomb lattice and $\ketbasic{\psi_{\rm up}^V}=\prod_{R}\ketbasic{\Up}_R$, which depends on $V$ in terms of bases written by $\ketbasic{\up}$ and $\ketbasic{\dw}$.
  Particularly, all $c_{RR'}^V$ take $+1$ in the subspace with $\sigma_j^z\sigma_k^z=+1$ for all $x$ bonds.
  We term this subspace $V_m$ hereafter.

  In the matrix representation on the basis of the direct product states $\ketbasic{\Ra}_R$ and $\ketbasic{\La}_R$, all matrix elements of $\tilde{\cal H}_{\rm eff}^{V_m}$ are negative.
  Using the relation $\sum_{ij}v_i^{*} A_{ij}v_j / |\bm{v}|^2 \geq \sum_{ij}|v_i| (-|A_{ij}|) |v_j| / |\bm{v}|^2$ for an arbitrary Hermitian matrix $A$ and vector $\bm{v}$, we find that the ground state energy of $\tilde{\cal H}_{\rm eff}^{V_m}$ is the smallest.
  Moreover, $\tilde{E}^V$, which is constant in each subspace, is the smallest for the subspace $V_m$.
  Thus, the ground state is always in the subspace $V_m$.

  Finally we obtain the two-dimensional transverse-field Ising model after the projection to the subspace $V_m$,
\begin{align}
{\cal H}_{\rm eff}^{V_m}=&-J_I\sum_{\means{RR'}_y} \tilde{\tau}_R^x \tilde{\tau}_{R'}^x
-(J_z+J_I) \sum_{\means{RR'}_z} \tilde{\tau}_R^x \tilde{\tau}_{R'}^x
 \nonumber\\
&-J_x \sum_{R} \tilde{\tau}_R^z-\frac{J_I N}{2}.
\end{align}
  This model undergoes a second-order phase transition between the ferromagnetic phase $\means{\tilde{\tau}^x}\ne 0$ and the quantum paramagnet with  $\means{\tilde{\tau}^z}= 0$, which corresponds to the spin-nematic liquid.
  Indeed, our MF results show a continuous transition, as shown in Fig.~1(c) and Figs.~2(e)--2(f) of the main text.
  Particularly, the case with $J_y = J_z = 0$ is mapped to the quantum Ising model on the perfect square lattice, in which case the quantum critical point is at $J_I=1/3.044(1)$~\cite{Rieger1999}, as indicated by the small arrow on the vertical axis in Fig.~1(c) of the main text.
  The result is consistent with our ED results.
  We therefore conclude that the transition from the spin nematic to ferromagnetic phase is of second order.

%% file: main3.0.bbl
%merlin.mbs apsrev4-1.bst 2010-07-25 4.21a (PWD, AO, DPC) hacked
%Control: key (0)
%Control: author (8) initials jnrlst
%Control: editor formatted (1) identically to author
%Control: production of article title (-1) disabled
%Control: page (0) single
%Control: year (1) truncated
%Control: production of eprint (0) enabled
\begin{thebibliography}{62}%
\makeatletter
\providecommand \@ifxundefined [1]{%
 \@ifx{#1\undefined}
}%
\providecommand \@ifnum [1]{%
 \ifnum #1\expandafter \@firstoftwo
 \else \expandafter \@secondoftwo
 \fi
}%
\providecommand \@ifx [1]{%
 \ifx #1\expandafter \@firstoftwo
 \else \expandafter \@secondoftwo
 \fi
}%
\providecommand \natexlab [1]{#1}%
\providecommand \enquote  [1]{``#1''}%
\providecommand \bibnamefont  [1]{#1}%
\providecommand \bibfnamefont [1]{#1}%
\providecommand \citenamefont [1]{#1}%
\providecommand \href@noop [0]{\@secondoftwo}%
\providecommand \href [0]{\begingroup \@sanitize@url \@href}%
\providecommand \@href[1]{\@@startlink{#1}\@@href}%
\providecommand \@@href[1]{\endgroup#1\@@endlink}%
\providecommand \@sanitize@url [0]{\catcode `\\12\catcode `\$12\catcode
  `\&12\catcode `\#12\catcode `\^12\catcode `\_12\catcode `\%12\relax}%
\providecommand \@@startlink[1]{}%
\providecommand \@@endlink[0]{}%
\providecommand \url  [0]{\begingroup\@sanitize@url \@url }%
\providecommand \@url [1]{\endgroup\@href {#1}{\urlprefix }}%
\providecommand \urlprefix  [0]{URL }%
\providecommand \Eprint [0]{\href }%
\providecommand \doibase [0]{http://dx.doi.org/}%
\providecommand \selectlanguage [0]{\@gobble}%
\providecommand \bibinfo  [0]{\@secondoftwo}%
\providecommand \bibfield  [0]{\@secondoftwo}%
\providecommand \translation [1]{[#1]}%
\providecommand \BibitemOpen [0]{}%
\providecommand \bibitemStop [0]{}%
\providecommand \bibitemNoStop [0]{.\EOS\space}%
\providecommand \EOS [0]{\spacefactor3000\relax}%
\providecommand \BibitemShut  [1]{\csname bibitem#1\endcsname}%
\let\auto@bib@innerbib\@empty
%</preamble>
\bibitem [{\citenamefont {Anderson}(1973)}]{Anderson1973153}%
  \BibitemOpen
  \bibfield  {author} {\bibinfo {author} {\bibfnamefont {P.}~\bibnamefont
  {Anderson}},\ }\href {\doibase
  http://dx.doi.org/10.1016/0025-5408(73)90167-0} {\bibfield  {journal}
  {\bibinfo  {journal} {Mater. Res. Bull.}\ }\textbf {\bibinfo {volume} {8}},\
  \bibinfo {pages} {153 } (\bibinfo {year} {1973})}\BibitemShut {NoStop}%
\bibitem [{\citenamefont {Shimizu}\ \emph {et~al.}(2003)\citenamefont
  {Shimizu}, \citenamefont {Miyagawa}, \citenamefont {Kanoda}, \citenamefont
  {Maesato},\ and\ \citenamefont {Saito}}]{PhysRevLett.91.107001}%
  \BibitemOpen
  \bibfield  {author} {\bibinfo {author} {\bibfnamefont {Y.}~\bibnamefont
  {Shimizu}}, \bibinfo {author} {\bibfnamefont {K.}~\bibnamefont {Miyagawa}},
  \bibinfo {author} {\bibfnamefont {K.}~\bibnamefont {Kanoda}}, \bibinfo
  {author} {\bibfnamefont {M.}~\bibnamefont {Maesato}}, \ and\ \bibinfo
  {author} {\bibfnamefont {G.}~\bibnamefont {Saito}},\ }\href {\doibase
  10.1103/PhysRevLett.91.107001} {\bibfield  {journal} {\bibinfo  {journal}
  {Phys. Rev. Lett.}\ }\textbf {\bibinfo {volume} {91}},\ \bibinfo {pages}
  {107001} (\bibinfo {year} {2003})}\BibitemShut {NoStop}%
\bibitem [{\citenamefont {Yamashita}\ \emph {et~al.}(2010)\citenamefont
  {Yamashita}, \citenamefont {Nakata}, \citenamefont {Senshu}, \citenamefont
  {Nagata}, \citenamefont {Yamamoto}, \citenamefont {Kato}, \citenamefont
  {Shibauchi},\ and\ \citenamefont {Matsuda}}]{ISI:000278318600025}%
  \BibitemOpen
  \bibfield  {author} {\bibinfo {author} {\bibfnamefont {M.}~\bibnamefont
  {Yamashita}}, \bibinfo {author} {\bibfnamefont {N.}~\bibnamefont {Nakata}},
  \bibinfo {author} {\bibfnamefont {Y.}~\bibnamefont {Senshu}}, \bibinfo
  {author} {\bibfnamefont {M.}~\bibnamefont {Nagata}}, \bibinfo {author}
  {\bibfnamefont {H.~M.}\ \bibnamefont {Yamamoto}}, \bibinfo {author}
  {\bibfnamefont {R.}~\bibnamefont {Kato}}, \bibinfo {author} {\bibfnamefont
  {T.}~\bibnamefont {Shibauchi}}, \ and\ \bibinfo {author} {\bibfnamefont
  {Y.}~\bibnamefont {Matsuda}},\ }\href {\doibase 10.1126/science.1188200}
  {\bibfield  {journal} {\bibinfo  {journal} {Science}\ }\textbf {\bibinfo
  {volume} {328}},\ \bibinfo {pages} {1246} (\bibinfo {year}
  {2010})}\BibitemShut {NoStop}%
\bibitem [{\citenamefont {Isono}\ \emph {et~al.}(2014)\citenamefont {Isono},
  \citenamefont {Kamo}, \citenamefont {Ueda}, \citenamefont {Takahashi},
  \citenamefont {Kimata}, \citenamefont {Tajima}, \citenamefont {Tsuchiya},
  \citenamefont {Terashima}, \citenamefont {Uji},\ and\ \citenamefont
  {Mori}}]{PhysRevLett.112.177201}%
  \BibitemOpen
  \bibfield  {author} {\bibinfo {author} {\bibfnamefont {T.}~\bibnamefont
  {Isono}}, \bibinfo {author} {\bibfnamefont {H.}~\bibnamefont {Kamo}},
  \bibinfo {author} {\bibfnamefont {A.}~\bibnamefont {Ueda}}, \bibinfo {author}
  {\bibfnamefont {K.}~\bibnamefont {Takahashi}}, \bibinfo {author}
  {\bibfnamefont {M.}~\bibnamefont {Kimata}}, \bibinfo {author} {\bibfnamefont
  {H.}~\bibnamefont {Tajima}}, \bibinfo {author} {\bibfnamefont
  {S.}~\bibnamefont {Tsuchiya}}, \bibinfo {author} {\bibfnamefont
  {T.}~\bibnamefont {Terashima}}, \bibinfo {author} {\bibfnamefont
  {S.}~\bibnamefont {Uji}}, \ and\ \bibinfo {author} {\bibfnamefont
  {H.}~\bibnamefont {Mori}},\ }\href {\doibase 10.1103/PhysRevLett.112.177201}
  {\bibfield  {journal} {\bibinfo  {journal} {Phys. Rev. Lett.}\ }\textbf
  {\bibinfo {volume} {112}},\ \bibinfo {pages} {177201} (\bibinfo {year}
  {2014})}\BibitemShut {NoStop}%
\bibitem [{\citenamefont {Nakatsuji}\ \emph {et~al.}(2005)\citenamefont
  {Nakatsuji}, \citenamefont {Nambu}, \citenamefont {Tonomura}, \citenamefont
  {Sakai}, \citenamefont {Jonas}, \citenamefont {Broholm}, \citenamefont
  {Tsunetsugu}, \citenamefont {Qiu},\ and\ \citenamefont
  {Maeno}}]{ISI:000231836700036}%
  \BibitemOpen
  \bibfield  {author} {\bibinfo {author} {\bibfnamefont {S.}~\bibnamefont
  {Nakatsuji}}, \bibinfo {author} {\bibfnamefont {Y.}~\bibnamefont {Nambu}},
  \bibinfo {author} {\bibfnamefont {H.}~\bibnamefont {Tonomura}}, \bibinfo
  {author} {\bibfnamefont {O.}~\bibnamefont {Sakai}}, \bibinfo {author}
  {\bibfnamefont {S.}~\bibnamefont {Jonas}}, \bibinfo {author} {\bibfnamefont
  {C.}~\bibnamefont {Broholm}}, \bibinfo {author} {\bibfnamefont
  {H.}~\bibnamefont {Tsunetsugu}}, \bibinfo {author} {\bibfnamefont
  {Y.}~\bibnamefont {Qiu}}, \ and\ \bibinfo {author} {\bibfnamefont
  {Y.}~\bibnamefont {Maeno}},\ }\href {\doibase 10.1126/science.1114727}
  {\bibfield  {journal} {\bibinfo  {journal} {Science}\ }\textbf {\bibinfo
  {volume} {309}},\ \bibinfo {pages} {1697} (\bibinfo {year}
  {2005})}\BibitemShut {NoStop}%
\bibitem [{\citenamefont {Helton}\ \emph {et~al.}(2007)\citenamefont {Helton},
  \citenamefont {Matan}, \citenamefont {Shores}, \citenamefont {Nytko},
  \citenamefont {Bartlett}, \citenamefont {Yoshida}, \citenamefont {Takano},
  \citenamefont {Suslov}, \citenamefont {Qiu}, \citenamefont {Chung},
  \citenamefont {Nocera},\ and\ \citenamefont {Lee}}]{PhysRevLett.98.107204}%
  \BibitemOpen
  \bibfield  {author} {\bibinfo {author} {\bibfnamefont {J.~S.}\ \bibnamefont
  {Helton}}, \bibinfo {author} {\bibfnamefont {K.}~\bibnamefont {Matan}},
  \bibinfo {author} {\bibfnamefont {M.~P.}\ \bibnamefont {Shores}}, \bibinfo
  {author} {\bibfnamefont {E.~A.}\ \bibnamefont {Nytko}}, \bibinfo {author}
  {\bibfnamefont {B.~M.}\ \bibnamefont {Bartlett}}, \bibinfo {author}
  {\bibfnamefont {Y.}~\bibnamefont {Yoshida}}, \bibinfo {author} {\bibfnamefont
  {Y.}~\bibnamefont {Takano}}, \bibinfo {author} {\bibfnamefont
  {A.}~\bibnamefont {Suslov}}, \bibinfo {author} {\bibfnamefont
  {Y.}~\bibnamefont {Qiu}}, \bibinfo {author} {\bibfnamefont {J.-H.}\
  \bibnamefont {Chung}}, \bibinfo {author} {\bibfnamefont {D.~G.}\ \bibnamefont
  {Nocera}}, \ and\ \bibinfo {author} {\bibfnamefont {Y.~S.}\ \bibnamefont
  {Lee}},\ }\href {\doibase 10.1103/PhysRevLett.98.107204} {\bibfield
  {journal} {\bibinfo  {journal} {Phys. Rev. Lett.}\ }\textbf {\bibinfo
  {volume} {98}},\ \bibinfo {pages} {107204} (\bibinfo {year}
  {2007})}\BibitemShut {NoStop}%
\bibitem [{\citenamefont {Okamoto}\ \emph {et~al.}(2007)\citenamefont
  {Okamoto}, \citenamefont {Nohara}, \citenamefont {Aruga-Katori},\ and\
  \citenamefont {Takagi}}]{PhysRevLett.99.137207}%
  \BibitemOpen
  \bibfield  {author} {\bibinfo {author} {\bibfnamefont {Y.}~\bibnamefont
  {Okamoto}}, \bibinfo {author} {\bibfnamefont {M.}~\bibnamefont {Nohara}},
  \bibinfo {author} {\bibfnamefont {H.}~\bibnamefont {Aruga-Katori}}, \ and\
  \bibinfo {author} {\bibfnamefont {H.}~\bibnamefont {Takagi}},\ }\href
  {\doibase 10.1103/PhysRevLett.99.137207} {\bibfield  {journal} {\bibinfo
  {journal} {Phys. Rev. Lett.}\ }\textbf {\bibinfo {volume} {99}},\ \bibinfo
  {pages} {137207} (\bibinfo {year} {2007})}\BibitemShut {NoStop}%
\bibitem [{\citenamefont {Wen}(2004)}]{Wen2004}%
  \BibitemOpen
  \bibfield  {author} {\bibinfo {author} {\bibfnamefont {X.-G.}\ \bibnamefont
  {Wen}},\ }\href@noop {} {\emph {\bibinfo {title} {Quantum Field Theory of
  Many-Body Systems}}}\ (\bibinfo  {publisher} {Oxford University Press},\
  \bibinfo {address} {Oxford},\ \bibinfo {year} {2004})\BibitemShut {NoStop}%
\bibitem [{\citenamefont {Misguich}(2011)}]{Lacroix2011C16}%
  \BibitemOpen
  \bibfield  {author} {\bibinfo {author} {\bibfnamefont {G.}~\bibnamefont
  {Misguich}},\ }\enquote {\bibinfo {title} {Quantum spin liquids and
  fractionalization},}\ in\ \href@noop {} {\emph {\bibinfo {booktitle}
  {Introduction to Frustrated Magnetism}}},\ \bibinfo {editor} {edited by\
  \bibinfo {editor} {\bibfnamefont {C.}~\bibnamefont {Lacroix}}, \bibinfo
  {editor} {\bibfnamefont {P.}~\bibnamefont {Mendels}}, \ and\ \bibinfo
  {editor} {\bibfnamefont {F.}~\bibnamefont {Mila}}}\ (\bibinfo  {publisher}
  {Springer},\ \bibinfo {address} {Heidelberg},\ \bibinfo {year} {2011})\
  Chap.~\bibinfo {chapter} {16}\BibitemShut {NoStop}%
\bibitem [{\citenamefont {Oshikawa}\ and\ \citenamefont
  {Senthil}(2006)}]{PhysRevLett.96.060601}%
  \BibitemOpen
  \bibfield  {author} {\bibinfo {author} {\bibfnamefont {M.}~\bibnamefont
  {Oshikawa}}\ and\ \bibinfo {author} {\bibfnamefont {T.}~\bibnamefont
  {Senthil}},\ }\href {\doibase 10.1103/PhysRevLett.96.060601} {\bibfield
  {journal} {\bibinfo  {journal} {Phys. Rev. Lett.}\ }\textbf {\bibinfo
  {volume} {96}},\ \bibinfo {pages} {060601} (\bibinfo {year}
  {2006})}\BibitemShut {NoStop}%
\bibitem [{\citenamefont {Yamashita}\ \emph {et~al.}(2008)\citenamefont
  {Yamashita}, \citenamefont {Nakazawa}, \citenamefont {Oguni}, \citenamefont
  {Oshima}, \citenamefont {Nojiri}, \citenamefont {Shimizu}, \citenamefont
  {Miyagawa},\ and\ \citenamefont {Kanoda}}]{yamashita2008thermodynamic}%
  \BibitemOpen
  \bibfield  {author} {\bibinfo {author} {\bibfnamefont {S.}~\bibnamefont
  {Yamashita}}, \bibinfo {author} {\bibfnamefont {Y.}~\bibnamefont {Nakazawa}},
  \bibinfo {author} {\bibfnamefont {M.}~\bibnamefont {Oguni}}, \bibinfo
  {author} {\bibfnamefont {Y.}~\bibnamefont {Oshima}}, \bibinfo {author}
  {\bibfnamefont {H.}~\bibnamefont {Nojiri}}, \bibinfo {author} {\bibfnamefont
  {Y.}~\bibnamefont {Shimizu}}, \bibinfo {author} {\bibfnamefont
  {K.}~\bibnamefont {Miyagawa}}, \ and\ \bibinfo {author} {\bibfnamefont
  {K.}~\bibnamefont {Kanoda}},\ }\href@noop {} {\bibfield  {journal} {\bibinfo
  {journal} {Nature Physics}\ }\textbf {\bibinfo {volume} {4}},\ \bibinfo
  {pages} {459} (\bibinfo {year} {2008})}\BibitemShut {NoStop}%
\bibitem [{\citenamefont {Yamashita}\ \emph {et~al.}(2009)\citenamefont
  {Yamashita}, \citenamefont {Nakata}, \citenamefont {Kasahara}, \citenamefont
  {Sasaki}, \citenamefont {Yoneyama}, \citenamefont {Kobayashi}, \citenamefont
  {Fujimoto}, \citenamefont {Shibauchi},\ and\ \citenamefont
  {Matsuda}}]{yamashita2009thermal}%
  \BibitemOpen
  \bibfield  {author} {\bibinfo {author} {\bibfnamefont {M.}~\bibnamefont
  {Yamashita}}, \bibinfo {author} {\bibfnamefont {N.}~\bibnamefont {Nakata}},
  \bibinfo {author} {\bibfnamefont {Y.}~\bibnamefont {Kasahara}}, \bibinfo
  {author} {\bibfnamefont {T.}~\bibnamefont {Sasaki}}, \bibinfo {author}
  {\bibfnamefont {N.}~\bibnamefont {Yoneyama}}, \bibinfo {author}
  {\bibfnamefont {N.}~\bibnamefont {Kobayashi}}, \bibinfo {author}
  {\bibfnamefont {S.}~\bibnamefont {Fujimoto}}, \bibinfo {author}
  {\bibfnamefont {T.}~\bibnamefont {Shibauchi}}, \ and\ \bibinfo {author}
  {\bibfnamefont {Y.}~\bibnamefont {Matsuda}},\ }\href {\doibase
  10.1038/nphys1134} {\bibfield  {journal} {\bibinfo  {journal} {Nat. Phys.}\
  }\textbf {\bibinfo {volume} {5}},\ \bibinfo {pages} {44} (\bibinfo {year}
  {2009})}\BibitemShut {NoStop}%
\bibitem [{\citenamefont {Kitaev}(2006)}]{Kitaev2006}%
  \BibitemOpen
  \bibfield  {author} {\bibinfo {author} {\bibfnamefont {A.}~\bibnamefont
  {Kitaev}},\ }\href {\doibase 10.1016/j.aop.2005.10.005} {\bibfield  {journal}
  {\bibinfo  {journal} {Ann. Phys. (N. Y.)}\ }\textbf {\bibinfo {volume}
  {321}},\ \bibinfo {pages} {2} (\bibinfo {year} {2006})}\BibitemShut {NoStop}%
\bibitem [{\citenamefont {Baskaran}\ \emph {et~al.}(2007)\citenamefont
  {Baskaran}, \citenamefont {Mandal},\ and\ \citenamefont
  {Shankar}}]{PhysRevLett.98.247201}%
  \BibitemOpen
  \bibfield  {author} {\bibinfo {author} {\bibfnamefont {G.}~\bibnamefont
  {Baskaran}}, \bibinfo {author} {\bibfnamefont {S.}~\bibnamefont {Mandal}}, \
  and\ \bibinfo {author} {\bibfnamefont {R.}~\bibnamefont {Shankar}},\ }\href
  {\doibase 10.1103/PhysRevLett.98.247201} {\bibfield  {journal} {\bibinfo
  {journal} {Phys. Rev. Lett.}\ }\textbf {\bibinfo {volume} {98}},\ \bibinfo
  {pages} {247201} (\bibinfo {year} {2007})}\BibitemShut {NoStop}%
\bibitem [{\citenamefont {Jackeli}\ and\ \citenamefont
  {Khaliullin}(2009)}]{PhysRevLett.102.017205}%
  \BibitemOpen
  \bibfield  {author} {\bibinfo {author} {\bibfnamefont {G.}~\bibnamefont
  {Jackeli}}\ and\ \bibinfo {author} {\bibfnamefont {G.}~\bibnamefont
  {Khaliullin}},\ }\href {\doibase 10.1103/PhysRevLett.102.017205} {\bibfield
  {journal} {\bibinfo  {journal} {Phys. Rev. Lett.}\ }\textbf {\bibinfo
  {volume} {102}},\ \bibinfo {pages} {017205} (\bibinfo {year}
  {2009})}\BibitemShut {NoStop}%
\bibitem [{\citenamefont {Knolle}\ \emph
  {et~al.}(2014{\natexlab{a}})\citenamefont {Knolle}, \citenamefont
  {Kovrizhin}, \citenamefont {Chalker},\ and\ \citenamefont
  {Moessner}}]{PhysRevLett.112.207203}%
  \BibitemOpen
  \bibfield  {author} {\bibinfo {author} {\bibfnamefont {J.}~\bibnamefont
  {Knolle}}, \bibinfo {author} {\bibfnamefont {D.~L.}\ \bibnamefont
  {Kovrizhin}}, \bibinfo {author} {\bibfnamefont {J.~T.}\ \bibnamefont
  {Chalker}}, \ and\ \bibinfo {author} {\bibfnamefont {R.}~\bibnamefont
  {Moessner}},\ }\href {\doibase 10.1103/PhysRevLett.112.207203} {\bibfield
  {journal} {\bibinfo  {journal} {Phys. Rev. Lett.}\ }\textbf {\bibinfo
  {volume} {112}},\ \bibinfo {pages} {207203} (\bibinfo {year}
  {2014}{\natexlab{a}})}\BibitemShut {NoStop}%
\bibitem [{\citenamefont {Knolle}\ \emph
  {et~al.}(2014{\natexlab{b}})\citenamefont {Knolle}, \citenamefont {Chern},
  \citenamefont {Kovrizhin}, \citenamefont {Moessner},\ and\ \citenamefont
  {Perkins}}]{PhysRevLett.113.187201}%
  \BibitemOpen
  \bibfield  {author} {\bibinfo {author} {\bibfnamefont {J.}~\bibnamefont
  {Knolle}}, \bibinfo {author} {\bibfnamefont {G.-W.}\ \bibnamefont {Chern}},
  \bibinfo {author} {\bibfnamefont {D.~L.}\ \bibnamefont {Kovrizhin}}, \bibinfo
  {author} {\bibfnamefont {R.}~\bibnamefont {Moessner}}, \ and\ \bibinfo
  {author} {\bibfnamefont {N.~B.}\ \bibnamefont {Perkins}},\ }\href {\doibase
  10.1103/PhysRevLett.113.187201} {\bibfield  {journal} {\bibinfo  {journal}
  {Phys. Rev. Lett.}\ }\textbf {\bibinfo {volume} {113}},\ \bibinfo {pages}
  {187201} (\bibinfo {year} {2014}{\natexlab{b}})}\BibitemShut {NoStop}%
\bibitem [{\citenamefont {Perreault}\ \emph {et~al.}(2015)\citenamefont
  {Perreault}, \citenamefont {Knolle}, \citenamefont {Perkins},\ and\
  \citenamefont {Burnell}}]{PhysRevB.92.094439}%
  \BibitemOpen
  \bibfield  {author} {\bibinfo {author} {\bibfnamefont {B.}~\bibnamefont
  {Perreault}}, \bibinfo {author} {\bibfnamefont {J.}~\bibnamefont {Knolle}},
  \bibinfo {author} {\bibfnamefont {N.~B.}\ \bibnamefont {Perkins}}, \ and\
  \bibinfo {author} {\bibfnamefont {F.~J.}\ \bibnamefont {Burnell}},\ }\href
  {\doibase 10.1103/PhysRevB.92.094439} {\bibfield  {journal} {\bibinfo
  {journal} {Phys. Rev. B}\ }\textbf {\bibinfo {volume} {92}},\ \bibinfo
  {pages} {094439} (\bibinfo {year} {2015})}\BibitemShut {NoStop}%
\bibitem [{\citenamefont {Knolle}\ \emph {et~al.}(2015)\citenamefont {Knolle},
  \citenamefont {Kovrizhin}, \citenamefont {Chalker},\ and\ \citenamefont
  {Moessner}}]{PhysRevB.92.115127}%
  \BibitemOpen
  \bibfield  {author} {\bibinfo {author} {\bibfnamefont {J.}~\bibnamefont
  {Knolle}}, \bibinfo {author} {\bibfnamefont {D.~L.}\ \bibnamefont
  {Kovrizhin}}, \bibinfo {author} {\bibfnamefont {J.~T.}\ \bibnamefont
  {Chalker}}, \ and\ \bibinfo {author} {\bibfnamefont {R.}~\bibnamefont
  {Moessner}},\ }\href {\doibase 10.1103/PhysRevB.92.115127} {\bibfield
  {journal} {\bibinfo  {journal} {Phys. Rev. B}\ }\textbf {\bibinfo {volume}
  {92}},\ \bibinfo {pages} {115127} (\bibinfo {year} {2015})}\BibitemShut
  {NoStop}%
\bibitem [{\citenamefont {Nasu}\ \emph {et~al.}(2014)\citenamefont {Nasu},
  \citenamefont {Udagawa},\ and\ \citenamefont
  {Motome}}]{PhysRevLett.113.197205}%
  \BibitemOpen
  \bibfield  {author} {\bibinfo {author} {\bibfnamefont {J.}~\bibnamefont
  {Nasu}}, \bibinfo {author} {\bibfnamefont {M.}~\bibnamefont {Udagawa}}, \
  and\ \bibinfo {author} {\bibfnamefont {Y.}~\bibnamefont {Motome}},\ }\href
  {\doibase 10.1103/PhysRevLett.113.197205} {\bibfield  {journal} {\bibinfo
  {journal} {Phys. Rev. Lett.}\ }\textbf {\bibinfo {volume} {113}},\ \bibinfo
  {pages} {197205} (\bibinfo {year} {2014})}\BibitemShut {NoStop}%
\bibitem [{\citenamefont {Nasu}\ \emph {et~al.}(2015)\citenamefont {Nasu},
  \citenamefont {Udagawa},\ and\ \citenamefont {Motome}}]{PhysRevB.92.115122}%
  \BibitemOpen
  \bibfield  {author} {\bibinfo {author} {\bibfnamefont {J.}~\bibnamefont
  {Nasu}}, \bibinfo {author} {\bibfnamefont {M.}~\bibnamefont {Udagawa}}, \
  and\ \bibinfo {author} {\bibfnamefont {Y.}~\bibnamefont {Motome}},\ }\href
  {\doibase 10.1103/PhysRevB.92.115122} {\bibfield  {journal} {\bibinfo
  {journal} {Phys. Rev. B}\ }\textbf {\bibinfo {volume} {92}},\ \bibinfo
  {pages} {115122} (\bibinfo {year} {2015})}\BibitemShut {NoStop}%
\bibitem [{\citenamefont {Nasu}\ and\ \citenamefont
  {Motome}(2015)}]{PhysRevLett.115.087203}%
  \BibitemOpen
  \bibfield  {author} {\bibinfo {author} {\bibfnamefont {J.}~\bibnamefont
  {Nasu}}\ and\ \bibinfo {author} {\bibfnamefont {Y.}~\bibnamefont {Motome}},\
  }\href {\doibase 10.1103/PhysRevLett.115.087203} {\bibfield  {journal}
  {\bibinfo  {journal} {Phys. Rev. Lett.}\ }\textbf {\bibinfo {volume} {115}},\
  \bibinfo {pages} {087203} (\bibinfo {year} {2015})}\BibitemShut {NoStop}%
\bibitem [{\citenamefont {Yoshitake}\ \emph {et~al.}(2016)\citenamefont
  {Yoshitake}, \citenamefont {Nasu},\ and\ \citenamefont
  {Motome}}]{yoshitake2016}%
  \BibitemOpen
  \bibfield  {author} {\bibinfo {author} {\bibfnamefont {J.}~\bibnamefont
  {Yoshitake}}, \bibinfo {author} {\bibfnamefont {J.}~\bibnamefont {Nasu}}, \
  and\ \bibinfo {author} {\bibfnamefont {Y.}~\bibnamefont {Motome}},\ }\href
  {\doibase 10.1103/PhysRevLett.117.157203} {\bibfield  {journal} {\bibinfo
  {journal} {Phys. Rev. Lett.}\ }\textbf {\bibinfo {volume} {117}},\ \bibinfo
  {pages} {157203} (\bibinfo {year} {2016})}\BibitemShut {NoStop}%
\bibitem [{\citenamefont {Nasu}\ \emph {et~al.}(2016)\citenamefont {Nasu},
  \citenamefont {Knolle}, \citenamefont {Kovrizhin}, \citenamefont {Motome},\
  and\ \citenamefont {Moessner}}]{Nasu2016nphys}%
  \BibitemOpen
  \bibfield  {author} {\bibinfo {author} {\bibfnamefont {J.}~\bibnamefont
  {Nasu}}, \bibinfo {author} {\bibfnamefont {J.}~\bibnamefont {Knolle}},
  \bibinfo {author} {\bibfnamefont {D.~L.}\ \bibnamefont {Kovrizhin}}, \bibinfo
  {author} {\bibfnamefont {Y.}~\bibnamefont {Motome}}, \ and\ \bibinfo {author}
  {\bibfnamefont {R.}~\bibnamefont {Moessner}},\ }\href {\doibase
  10.1038/nphys3809} {\bibfield  {journal} {\bibinfo  {journal} {Nat. Phys.}\
  }\textbf {\bibinfo {volume} {12}},\ \bibinfo {pages} {912} (\bibinfo {year}
  {2016})}\BibitemShut {NoStop}%
\bibitem [{\citenamefont {Singh}\ \emph {et~al.}(2012)\citenamefont {Singh},
  \citenamefont {Manni}, \citenamefont {Reuther}, \citenamefont {Berlijn},
  \citenamefont {Thomale}, \citenamefont {Ku}, \citenamefont {Trebst},\ and\
  \citenamefont {Gegenwart}}]{PhysRevLett.108.127203}%
  \BibitemOpen
  \bibfield  {author} {\bibinfo {author} {\bibfnamefont {Y.}~\bibnamefont
  {Singh}}, \bibinfo {author} {\bibfnamefont {S.}~\bibnamefont {Manni}},
  \bibinfo {author} {\bibfnamefont {J.}~\bibnamefont {Reuther}}, \bibinfo
  {author} {\bibfnamefont {T.}~\bibnamefont {Berlijn}}, \bibinfo {author}
  {\bibfnamefont {R.}~\bibnamefont {Thomale}}, \bibinfo {author} {\bibfnamefont
  {W.}~\bibnamefont {Ku}}, \bibinfo {author} {\bibfnamefont {S.}~\bibnamefont
  {Trebst}}, \ and\ \bibinfo {author} {\bibfnamefont {P.}~\bibnamefont
  {Gegenwart}},\ }\href {\doibase 10.1103/PhysRevLett.108.127203} {\bibfield
  {journal} {\bibinfo  {journal} {Phys. Rev. Lett.}\ }\textbf {\bibinfo
  {volume} {108}},\ \bibinfo {pages} {127203} (\bibinfo {year}
  {2012})}\BibitemShut {NoStop}%
\bibitem [{\citenamefont {Comin}\ \emph {et~al.}(2012)\citenamefont {Comin},
  \citenamefont {Levy}, \citenamefont {Ludbrook}, \citenamefont {Zhu},
  \citenamefont {Veenstra}, \citenamefont {Rosen}, \citenamefont {Singh},
  \citenamefont {Gegenwart}, \citenamefont {Stricker}, \citenamefont {Hancock},
  \citenamefont {van~der Marel}, \citenamefont {Elfimov},\ and\ \citenamefont
  {Damascelli}}]{PhysRevLett.109.266406}%
  \BibitemOpen
  \bibfield  {author} {\bibinfo {author} {\bibfnamefont {R.}~\bibnamefont
  {Comin}}, \bibinfo {author} {\bibfnamefont {G.}~\bibnamefont {Levy}},
  \bibinfo {author} {\bibfnamefont {B.}~\bibnamefont {Ludbrook}}, \bibinfo
  {author} {\bibfnamefont {Z.-H.}\ \bibnamefont {Zhu}}, \bibinfo {author}
  {\bibfnamefont {C.~N.}\ \bibnamefont {Veenstra}}, \bibinfo {author}
  {\bibfnamefont {J.~A.}\ \bibnamefont {Rosen}}, \bibinfo {author}
  {\bibfnamefont {Y.}~\bibnamefont {Singh}}, \bibinfo {author} {\bibfnamefont
  {P.}~\bibnamefont {Gegenwart}}, \bibinfo {author} {\bibfnamefont
  {D.}~\bibnamefont {Stricker}}, \bibinfo {author} {\bibfnamefont {J.~N.}\
  \bibnamefont {Hancock}}, \bibinfo {author} {\bibfnamefont {D.}~\bibnamefont
  {van~der Marel}}, \bibinfo {author} {\bibfnamefont {I.~S.}\ \bibnamefont
  {Elfimov}}, \ and\ \bibinfo {author} {\bibfnamefont {A.}~\bibnamefont
  {Damascelli}},\ }\href {\doibase 10.1103/PhysRevLett.109.266406} {\bibfield
  {journal} {\bibinfo  {journal} {Phys. Rev. Lett.}\ }\textbf {\bibinfo
  {volume} {109}},\ \bibinfo {pages} {266406} (\bibinfo {year}
  {2012})}\BibitemShut {NoStop}%
\bibitem [{\citenamefont {Kubota}\ \emph {et~al.}(2015)\citenamefont {Kubota},
  \citenamefont {Tanaka}, \citenamefont {Ono}, \citenamefont {Narumi},\ and\
  \citenamefont {Kindo}}]{PhysRevB.91.094422}%
  \BibitemOpen
  \bibfield  {author} {\bibinfo {author} {\bibfnamefont {Y.}~\bibnamefont
  {Kubota}}, \bibinfo {author} {\bibfnamefont {H.}~\bibnamefont {Tanaka}},
  \bibinfo {author} {\bibfnamefont {T.}~\bibnamefont {Ono}}, \bibinfo {author}
  {\bibfnamefont {Y.}~\bibnamefont {Narumi}}, \ and\ \bibinfo {author}
  {\bibfnamefont {K.}~\bibnamefont {Kindo}},\ }\href {\doibase
  10.1103/PhysRevB.91.094422} {\bibfield  {journal} {\bibinfo  {journal} {Phys.
  Rev. B}\ }\textbf {\bibinfo {volume} {91}},\ \bibinfo {pages} {094422}
  (\bibinfo {year} {2015})}\BibitemShut {NoStop}%
\bibitem [{\citenamefont {Plumb}\ \emph {et~al.}(2014)\citenamefont {Plumb},
  \citenamefont {Clancy}, \citenamefont {Sandilands}, \citenamefont {Shankar},
  \citenamefont {Hu}, \citenamefont {Burch}, \citenamefont {Kee},\ and\
  \citenamefont {Kim}}]{PhysRevB.90.041112}%
  \BibitemOpen
  \bibfield  {author} {\bibinfo {author} {\bibfnamefont {K.~W.}\ \bibnamefont
  {Plumb}}, \bibinfo {author} {\bibfnamefont {J.~P.}\ \bibnamefont {Clancy}},
  \bibinfo {author} {\bibfnamefont {L.~J.}\ \bibnamefont {Sandilands}},
  \bibinfo {author} {\bibfnamefont {V.~V.}\ \bibnamefont {Shankar}}, \bibinfo
  {author} {\bibfnamefont {Y.~F.}\ \bibnamefont {Hu}}, \bibinfo {author}
  {\bibfnamefont {K.~S.}\ \bibnamefont {Burch}}, \bibinfo {author}
  {\bibfnamefont {H.-Y.}\ \bibnamefont {Kee}}, \ and\ \bibinfo {author}
  {\bibfnamefont {Y.-J.}\ \bibnamefont {Kim}},\ }\href {\doibase
  10.1103/PhysRevB.90.041112} {\bibfield  {journal} {\bibinfo  {journal} {Phys.
  Rev. B}\ }\textbf {\bibinfo {volume} {90}},\ \bibinfo {pages} {041112}
  (\bibinfo {year} {2014})}\BibitemShut {NoStop}%
\bibitem [{\citenamefont {Chaloupka}\ \emph {et~al.}(2010)\citenamefont
  {Chaloupka}, \citenamefont {Jackeli},\ and\ \citenamefont
  {Khaliullin}}]{PhysRevLett.105.027204}%
  \BibitemOpen
  \bibfield  {author} {\bibinfo {author} {\bibfnamefont {J.}~\bibnamefont
  {Chaloupka}}, \bibinfo {author} {\bibfnamefont {G.}~\bibnamefont {Jackeli}},
  \ and\ \bibinfo {author} {\bibfnamefont {G.}~\bibnamefont {Khaliullin}},\
  }\href {\doibase 10.1103/PhysRevLett.105.027204} {\bibfield  {journal}
  {\bibinfo  {journal} {Phys. Rev. Lett.}\ }\textbf {\bibinfo {volume} {105}},\
  \bibinfo {pages} {027204} (\bibinfo {year} {2010})}\BibitemShut {NoStop}%
\bibitem [{\citenamefont {Chaloupka}\ \emph {et~al.}(2013)\citenamefont
  {Chaloupka}, \citenamefont {Jackeli},\ and\ \citenamefont
  {Khaliullin}}]{PhysRevLett.110.097204}%
  \BibitemOpen
  \bibfield  {author} {\bibinfo {author} {\bibfnamefont {J.}~\bibnamefont
  {Chaloupka}}, \bibinfo {author} {\bibfnamefont {G.}~\bibnamefont {Jackeli}},
  \ and\ \bibinfo {author} {\bibfnamefont {G.}~\bibnamefont {Khaliullin}},\
  }\href {\doibase 10.1103/PhysRevLett.110.097204} {\bibfield  {journal}
  {\bibinfo  {journal} {Phys. Rev. Lett.}\ }\textbf {\bibinfo {volume} {110}},\
  \bibinfo {pages} {097204} (\bibinfo {year} {2013})}\BibitemShut {NoStop}%
\bibitem [{\citenamefont {Yamaji}\ \emph {et~al.}(2014)\citenamefont {Yamaji},
  \citenamefont {Nomura}, \citenamefont {Kurita}, \citenamefont {Arita},\ and\
  \citenamefont {Imada}}]{PhysRevLett.113.107201}%
  \BibitemOpen
  \bibfield  {author} {\bibinfo {author} {\bibfnamefont {Y.}~\bibnamefont
  {Yamaji}}, \bibinfo {author} {\bibfnamefont {Y.}~\bibnamefont {Nomura}},
  \bibinfo {author} {\bibfnamefont {M.}~\bibnamefont {Kurita}}, \bibinfo
  {author} {\bibfnamefont {R.}~\bibnamefont {Arita}}, \ and\ \bibinfo {author}
  {\bibfnamefont {M.}~\bibnamefont {Imada}},\ }\href {\doibase
  10.1103/PhysRevLett.113.107201} {\bibfield  {journal} {\bibinfo  {journal}
  {Phys. Rev. Lett.}\ }\textbf {\bibinfo {volume} {113}},\ \bibinfo {pages}
  {107201} (\bibinfo {year} {2014})}\BibitemShut {NoStop}%
\bibitem [{\citenamefont {Reuther}\ \emph {et~al.}(2011)\citenamefont
  {Reuther}, \citenamefont {Thomale},\ and\ \citenamefont
  {Trebst}}]{PhysRevB.84.100406}%
  \BibitemOpen
  \bibfield  {author} {\bibinfo {author} {\bibfnamefont {J.}~\bibnamefont
  {Reuther}}, \bibinfo {author} {\bibfnamefont {R.}~\bibnamefont {Thomale}}, \
  and\ \bibinfo {author} {\bibfnamefont {S.}~\bibnamefont {Trebst}},\ }\href
  {\doibase 10.1103/PhysRevB.84.100406} {\bibfield  {journal} {\bibinfo
  {journal} {Phys. Rev. B}\ }\textbf {\bibinfo {volume} {84}},\ \bibinfo
  {pages} {100406} (\bibinfo {year} {2011})}\BibitemShut {NoStop}%
\bibitem [{\citenamefont {Kimchi}\ and\ \citenamefont
  {You}(2011)}]{PhysRevB.84.180407}%
  \BibitemOpen
  \bibfield  {author} {\bibinfo {author} {\bibfnamefont {I.}~\bibnamefont
  {Kimchi}}\ and\ \bibinfo {author} {\bibfnamefont {Y.-Z.}\ \bibnamefont
  {You}},\ }\href {\doibase 10.1103/PhysRevB.84.180407} {\bibfield  {journal}
  {\bibinfo  {journal} {Phys. Rev. B}\ }\textbf {\bibinfo {volume} {84}},\
  \bibinfo {pages} {180407} (\bibinfo {year} {2011})}\BibitemShut {NoStop}%
\bibitem [{\citenamefont {Katukuri}\ \emph {et~al.}(2014)\citenamefont
  {Katukuri}, \citenamefont {Nishimoto}, \citenamefont {Yushankhai},
  \citenamefont {Stoyanova}, \citenamefont {Kandpal}, \citenamefont {Choi},
  \citenamefont {Coldea}, \citenamefont {Rousochatzakis}, \citenamefont
  {Hozoi},\ and\ \citenamefont {van~den Brink}}]{1367-2630-16-1-013056}%
  \BibitemOpen
  \bibfield  {author} {\bibinfo {author} {\bibfnamefont {V.~M.}\ \bibnamefont
  {Katukuri}}, \bibinfo {author} {\bibfnamefont {S.}~\bibnamefont {Nishimoto}},
  \bibinfo {author} {\bibfnamefont {V.}~\bibnamefont {Yushankhai}}, \bibinfo
  {author} {\bibfnamefont {A.}~\bibnamefont {Stoyanova}}, \bibinfo {author}
  {\bibfnamefont {H.}~\bibnamefont {Kandpal}}, \bibinfo {author} {\bibfnamefont
  {S.}~\bibnamefont {Choi}}, \bibinfo {author} {\bibfnamefont {R.}~\bibnamefont
  {Coldea}}, \bibinfo {author} {\bibfnamefont {I.}~\bibnamefont
  {Rousochatzakis}}, \bibinfo {author} {\bibfnamefont {L.}~\bibnamefont
  {Hozoi}}, \ and\ \bibinfo {author} {\bibfnamefont {J.}~\bibnamefont {van~den
  Brink}},\ }\href {http://stacks.iop.org/1367-2630/16/i=1/a=013056} {\bibfield
   {journal} {\bibinfo  {journal} {New J. Phys.}\ }\textbf {\bibinfo {volume}
  {16}},\ \bibinfo {pages} {013056} (\bibinfo {year} {2014})}\BibitemShut
  {NoStop}%
\bibitem [{\citenamefont {Foyevtsova}\ \emph {et~al.}(2013)\citenamefont
  {Foyevtsova}, \citenamefont {Jeschke}, \citenamefont {Mazin}, \citenamefont
  {Khomskii},\ and\ \citenamefont {Valent\'{\i}}}]{PhysRevB.88.035107}%
  \BibitemOpen
  \bibfield  {author} {\bibinfo {author} {\bibfnamefont {K.}~\bibnamefont
  {Foyevtsova}}, \bibinfo {author} {\bibfnamefont {H.~O.}\ \bibnamefont
  {Jeschke}}, \bibinfo {author} {\bibfnamefont {I.~I.}\ \bibnamefont {Mazin}},
  \bibinfo {author} {\bibfnamefont {D.~I.}\ \bibnamefont {Khomskii}}, \ and\
  \bibinfo {author} {\bibfnamefont {R.}~\bibnamefont {Valent\'{\i}}},\ }\href
  {\doibase 10.1103/PhysRevB.88.035107} {\bibfield  {journal} {\bibinfo
  {journal} {Phys. Rev. B}\ }\textbf {\bibinfo {volume} {88}},\ \bibinfo
  {pages} {035107} (\bibinfo {year} {2013})}\BibitemShut {NoStop}%
\bibitem [{\citenamefont {Sizyuk}\ \emph {et~al.}(2014)\citenamefont {Sizyuk},
  \citenamefont {Price}, \citenamefont {W\"olfle},\ and\ \citenamefont
  {Perkins}}]{PhysRevB.90.155126}%
  \BibitemOpen
  \bibfield  {author} {\bibinfo {author} {\bibfnamefont {Y.}~\bibnamefont
  {Sizyuk}}, \bibinfo {author} {\bibfnamefont {C.}~\bibnamefont {Price}},
  \bibinfo {author} {\bibfnamefont {P.}~\bibnamefont {W\"olfle}}, \ and\
  \bibinfo {author} {\bibfnamefont {N.~B.}\ \bibnamefont {Perkins}},\ }\href
  {\doibase 10.1103/PhysRevB.90.155126} {\bibfield  {journal} {\bibinfo
  {journal} {Phys. Rev. B}\ }\textbf {\bibinfo {volume} {90}},\ \bibinfo
  {pages} {155126} (\bibinfo {year} {2014})}\BibitemShut {NoStop}%
\bibitem [{\citenamefont {Suzuki}\ \emph {et~al.}(2015)\citenamefont {Suzuki},
  \citenamefont {Yamada}, \citenamefont {Yamaji},\ and\ \citenamefont
  {Suga}}]{PhysRevB.92.184411}%
  \BibitemOpen
  \bibfield  {author} {\bibinfo {author} {\bibfnamefont {T.}~\bibnamefont
  {Suzuki}}, \bibinfo {author} {\bibfnamefont {T.}~\bibnamefont {Yamada}},
  \bibinfo {author} {\bibfnamefont {Y.}~\bibnamefont {Yamaji}}, \ and\ \bibinfo
  {author} {\bibfnamefont {S.}~\bibnamefont {Suga}},\ }\href {\doibase
  10.1103/PhysRevB.92.184411} {\bibfield  {journal} {\bibinfo  {journal} {Phys.
  Rev. B}\ }\textbf {\bibinfo {volume} {92}},\ \bibinfo {pages} {184411}
  (\bibinfo {year} {2015})}\BibitemShut {NoStop}%
\bibitem [{\citenamefont {Kim}\ and\ \citenamefont
  {Kee}(2016)}]{PhysRevB.93.155143}%
  \BibitemOpen
  \bibfield  {author} {\bibinfo {author} {\bibfnamefont {H.-S.}\ \bibnamefont
  {Kim}}\ and\ \bibinfo {author} {\bibfnamefont {H.-Y.}\ \bibnamefont {Kee}},\
  }\href {\doibase 10.1103/PhysRevB.93.155143} {\bibfield  {journal} {\bibinfo
  {journal} {Phys. Rev. B}\ }\textbf {\bibinfo {volume} {93}},\ \bibinfo
  {pages} {155143} (\bibinfo {year} {2016})}\BibitemShut {NoStop}%
\bibitem [{\citenamefont {Banerjee}\ \emph {et~al.}(2016)\citenamefont
  {Banerjee}, \citenamefont {Bridges}, \citenamefont {Yan}, \citenamefont
  {Aczel}, \citenamefont {Li}, \citenamefont {Stone}, \citenamefont {Granroth},
  \citenamefont {Lumsden}, \citenamefont {Yiu}, \citenamefont {Knolle} \emph
  {et~al.}}]{banerjee2016proximate}%
  \BibitemOpen
  \bibfield  {author} {\bibinfo {author} {\bibfnamefont {A.}~\bibnamefont
  {Banerjee}}, \bibinfo {author} {\bibfnamefont {C.}~\bibnamefont {Bridges}},
  \bibinfo {author} {\bibfnamefont {J.-Q.}\ \bibnamefont {Yan}}, \bibinfo
  {author} {\bibfnamefont {A.}~\bibnamefont {Aczel}}, \bibinfo {author}
  {\bibfnamefont {L.}~\bibnamefont {Li}}, \bibinfo {author} {\bibfnamefont
  {M.}~\bibnamefont {Stone}}, \bibinfo {author} {\bibfnamefont
  {G.}~\bibnamefont {Granroth}}, \bibinfo {author} {\bibfnamefont
  {M.}~\bibnamefont {Lumsden}}, \bibinfo {author} {\bibfnamefont
  {Y.}~\bibnamefont {Yiu}}, \bibinfo {author} {\bibfnamefont {J.}~\bibnamefont
  {Knolle}},  \emph {et~al.},\ }\href {\doibase 10.1038/nmat4604} {\bibfield
  {journal} {\bibinfo  {journal} {Nat. Mater.}\ }\textbf {\bibinfo {volume}
  {15}},\ \bibinfo {pages} {733} (\bibinfo {year} {2016})}\BibitemShut
  {NoStop}%
\bibitem [{\citenamefont {Yamaji}\ \emph {et~al.}(2016)\citenamefont {Yamaji},
  \citenamefont {Suzuki}, \citenamefont {Yamada}, \citenamefont {Suga},
  \citenamefont {Kawashima},\ and\ \citenamefont {Imada}}]{PhysRevB.93.174425}%
  \BibitemOpen
  \bibfield  {author} {\bibinfo {author} {\bibfnamefont {Y.}~\bibnamefont
  {Yamaji}}, \bibinfo {author} {\bibfnamefont {T.}~\bibnamefont {Suzuki}},
  \bibinfo {author} {\bibfnamefont {T.}~\bibnamefont {Yamada}}, \bibinfo
  {author} {\bibfnamefont {S.}~\bibnamefont {Suga}}, \bibinfo {author}
  {\bibfnamefont {N.}~\bibnamefont {Kawashima}}, \ and\ \bibinfo {author}
  {\bibfnamefont {M.}~\bibnamefont {Imada}},\ }\href {\doibase
  10.1103/PhysRevB.93.174425} {\bibfield  {journal} {\bibinfo  {journal} {Phys.
  Rev. B}\ }\textbf {\bibinfo {volume} {93}},\ \bibinfo {pages} {174425}
  (\bibinfo {year} {2016})}\BibitemShut {NoStop}%
\bibitem [{\citenamefont {Kamiya}\ \emph {et~al.}(2015)\citenamefont {Kamiya},
  \citenamefont {Kato}, \citenamefont {Nasu},\ and\ \citenamefont
  {Motome}}]{PhysRevB.92.100403}%
  \BibitemOpen
  \bibfield  {author} {\bibinfo {author} {\bibfnamefont {Y.}~\bibnamefont
  {Kamiya}}, \bibinfo {author} {\bibfnamefont {Y.}~\bibnamefont {Kato}},
  \bibinfo {author} {\bibfnamefont {J.}~\bibnamefont {Nasu}}, \ and\ \bibinfo
  {author} {\bibfnamefont {Y.}~\bibnamefont {Motome}},\ }\href {\doibase
  10.1103/PhysRevB.92.100403} {\bibfield  {journal} {\bibinfo  {journal} {Phys.
  Rev. B}\ }\textbf {\bibinfo {volume} {92}},\ \bibinfo {pages} {100403}
  (\bibinfo {year} {2015})}\BibitemShut {NoStop}%
\bibitem [{\citenamefont {Sandilands}\ \emph {et~al.}(2015)\citenamefont
  {Sandilands}, \citenamefont {Tian}, \citenamefont {Plumb}, \citenamefont
  {Kim},\ and\ \citenamefont {Burch}}]{PhysRevLett.114.147201}%
  \BibitemOpen
  \bibfield  {author} {\bibinfo {author} {\bibfnamefont {L.~J.}\ \bibnamefont
  {Sandilands}}, \bibinfo {author} {\bibfnamefont {Y.}~\bibnamefont {Tian}},
  \bibinfo {author} {\bibfnamefont {K.~W.}\ \bibnamefont {Plumb}}, \bibinfo
  {author} {\bibfnamefont {Y.-J.}\ \bibnamefont {Kim}}, \ and\ \bibinfo
  {author} {\bibfnamefont {K.~S.}\ \bibnamefont {Burch}},\ }\href {\doibase
  10.1103/PhysRevLett.114.147201} {\bibfield  {journal} {\bibinfo  {journal}
  {Phys. Rev. Lett.}\ }\textbf {\bibinfo {volume} {114}},\ \bibinfo {pages}
  {147201} (\bibinfo {year} {2015})}\BibitemShut {NoStop}%
\bibitem [{\citenamefont {Mandal}\ \emph {et~al.}(2011)\citenamefont {Mandal},
  \citenamefont {Bhattacharjee}, \citenamefont {Sengupta}, \citenamefont
  {Shankar},\ and\ \citenamefont {Baskaran}}]{PhysRevB.84.155121}%
  \BibitemOpen
  \bibfield  {author} {\bibinfo {author} {\bibfnamefont {S.}~\bibnamefont
  {Mandal}}, \bibinfo {author} {\bibfnamefont {S.}~\bibnamefont
  {Bhattacharjee}}, \bibinfo {author} {\bibfnamefont {K.}~\bibnamefont
  {Sengupta}}, \bibinfo {author} {\bibfnamefont {R.}~\bibnamefont {Shankar}}, \
  and\ \bibinfo {author} {\bibfnamefont {G.}~\bibnamefont {Baskaran}},\ }\href
  {\doibase 10.1103/PhysRevB.84.155121} {\bibfield  {journal} {\bibinfo
  {journal} {Phys. Rev. B}\ }\textbf {\bibinfo {volume} {84}},\ \bibinfo
  {pages} {155121} (\bibinfo {year} {2011})}\BibitemShut {NoStop}%
\bibitem [{\citenamefont {Sugiura}\ and\ \citenamefont
  {Shimizu}(2012)}]{PhysRevLett.108.240401}%
  \BibitemOpen
  \bibfield  {author} {\bibinfo {author} {\bibfnamefont {S.}~\bibnamefont
  {Sugiura}}\ and\ \bibinfo {author} {\bibfnamefont {A.}~\bibnamefont
  {Shimizu}},\ }\href {\doibase 10.1103/PhysRevLett.108.240401} {\bibfield
  {journal} {\bibinfo  {journal} {Phys. Rev. Lett.}\ }\textbf {\bibinfo
  {volume} {108}},\ \bibinfo {pages} {240401} (\bibinfo {year}
  {2012})}\BibitemShut {NoStop}%
\bibitem [{\citenamefont {Sugiura}\ and\ \citenamefont
  {Shimizu}(2013)}]{PhysRevLett.111.010401}%
  \BibitemOpen
  \bibfield  {author} {\bibinfo {author} {\bibfnamefont {S.}~\bibnamefont
  {Sugiura}}\ and\ \bibinfo {author} {\bibfnamefont {A.}~\bibnamefont
  {Shimizu}},\ }\href {\doibase 10.1103/PhysRevLett.111.010401} {\bibfield
  {journal} {\bibinfo  {journal} {Phys. Rev. Lett.}\ }\textbf {\bibinfo
  {volume} {111}},\ \bibinfo {pages} {010401} (\bibinfo {year}
  {2013})}\BibitemShut {NoStop}%
\bibitem [{\citenamefont {Hyuga}\ \emph {et~al.}(2014)\citenamefont {Hyuga},
  \citenamefont {Sugiura}, \citenamefont {Sakai},\ and\ \citenamefont
  {Shimizu}}]{PhysRevB.90.121110}%
  \BibitemOpen
  \bibfield  {author} {\bibinfo {author} {\bibfnamefont {M.}~\bibnamefont
  {Hyuga}}, \bibinfo {author} {\bibfnamefont {S.}~\bibnamefont {Sugiura}},
  \bibinfo {author} {\bibfnamefont {K.}~\bibnamefont {Sakai}}, \ and\ \bibinfo
  {author} {\bibfnamefont {A.}~\bibnamefont {Shimizu}},\ }\href {\doibase
  10.1103/PhysRevB.90.121110} {\bibfield  {journal} {\bibinfo  {journal} {Phys.
  Rev. B}\ }\textbf {\bibinfo {volume} {90}},\ \bibinfo {pages} {121110}
  (\bibinfo {year} {2014})}\BibitemShut {NoStop}%
\bibitem [{\citenamefont {Chen}\ and\ \citenamefont
  {Hu}(2007)}]{PhysRevB.76.193101}%
  \BibitemOpen
  \bibfield  {author} {\bibinfo {author} {\bibfnamefont {H.-D.}\ \bibnamefont
  {Chen}}\ and\ \bibinfo {author} {\bibfnamefont {J.}~\bibnamefont {Hu}},\
  }\href {\doibase 10.1103/PhysRevB.76.193101} {\bibfield  {journal} {\bibinfo
  {journal} {Phys. Rev. B}\ }\textbf {\bibinfo {volume} {76}},\ \bibinfo
  {pages} {193101} (\bibinfo {year} {2007})}\BibitemShut {NoStop}%
\bibitem [{\citenamefont {Feng}\ \emph {et~al.}(2007)\citenamefont {Feng},
  \citenamefont {Zhang},\ and\ \citenamefont {Xiang}}]{PhysRevLett.98.087204}%
  \BibitemOpen
  \bibfield  {author} {\bibinfo {author} {\bibfnamefont {X.-Y.}\ \bibnamefont
  {Feng}}, \bibinfo {author} {\bibfnamefont {G.-M.}\ \bibnamefont {Zhang}}, \
  and\ \bibinfo {author} {\bibfnamefont {T.}~\bibnamefont {Xiang}},\ }\href
  {\doibase 10.1103/PhysRevLett.98.087204} {\bibfield  {journal} {\bibinfo
  {journal} {Phys. Rev. Lett.}\ }\textbf {\bibinfo {volume} {98}},\ \bibinfo
  {pages} {087204} (\bibinfo {year} {2007})}\BibitemShut {NoStop}%
\bibitem [{\citenamefont {Chen}\ and\ \citenamefont
  {Nussinov}(2008)}]{1751-8121-41-7-075001}%
  \BibitemOpen
  \bibfield  {author} {\bibinfo {author} {\bibfnamefont {H.-D.}\ \bibnamefont
  {Chen}}\ and\ \bibinfo {author} {\bibfnamefont {Z.}~\bibnamefont
  {Nussinov}},\ }\href {http://stacks.iop.org/1751-8121/41/i=7/a=075001}
  {\bibfield  {journal} {\bibinfo  {journal} {J. Phys. A}\ }\textbf {\bibinfo
  {volume} {41}},\ \bibinfo {pages} {075001} (\bibinfo {year}
  {2008})}\BibitemShut {NoStop}%
\bibitem [{Note1()}]{Note1}%
  \BibitemOpen
  \bibinfo {note} {The gapped-gapless boundary for $J_I > 0$ seems almost
  insensitive to $J_I$ according to our MF calculation, while the ED for 24
  spins is not efficient to clarify this subtlety.}\BibitemShut {Stop}%
\bibitem [{\citenamefont {Penc}\ and\ \citenamefont
  {L\"auchli}(2011)}]{Lacroix2011C13}%
  \BibitemOpen
  \bibfield  {author} {\bibinfo {author} {\bibfnamefont {K.}~\bibnamefont
  {Penc}}\ and\ \bibinfo {author} {\bibfnamefont {A.}~\bibnamefont
  {L\"auchli}},\ }\enquote {\bibinfo {title} {Spin nematic phases in quantum
  spin systems},}\ in\ \href@noop {} {\emph {\bibinfo {booktitle} {Introduction
  to Frustrated Magnetism}}},\ \bibinfo {editor} {edited by\ \bibinfo {editor}
  {\bibfnamefont {C.}~\bibnamefont {Lacroix}}, \bibinfo {editor} {\bibfnamefont
  {P.}~\bibnamefont {Mendels}}, \ and\ \bibinfo {editor} {\bibfnamefont
  {F.}~\bibnamefont {Mila}}}\ (\bibinfo  {publisher} {Springer},\ \bibinfo
  {address} {Heidelberg},\ \bibinfo {year} {2011})\ Chap.~\bibinfo {chapter}
  {13}\BibitemShut {NoStop}%
\bibitem [{sup()}]{suppl}%
  \BibitemOpen
  \href@noop {} {}\bibinfo {note} {See Supplemental Material}\BibitemShut
  {NoStop}%
\bibitem [{\citenamefont {Vidal}\ \emph {et~al.}(2009)\citenamefont {Vidal},
  \citenamefont {Thomale}, \citenamefont {Schmidt},\ and\ \citenamefont
  {Dusuel}}]{PhysRevB.80.081104}%
  \BibitemOpen
  \bibfield  {author} {\bibinfo {author} {\bibfnamefont {J.}~\bibnamefont
  {Vidal}}, \bibinfo {author} {\bibfnamefont {R.}~\bibnamefont {Thomale}},
  \bibinfo {author} {\bibfnamefont {K.~P.}\ \bibnamefont {Schmidt}}, \ and\
  \bibinfo {author} {\bibfnamefont {S.}~\bibnamefont {Dusuel}},\ }\href
  {\doibase 10.1103/PhysRevB.80.081104} {\bibfield  {journal} {\bibinfo
  {journal} {Phys. Rev. B}\ }\textbf {\bibinfo {volume} {80}},\ \bibinfo
  {pages} {081104} (\bibinfo {year} {2009})}\BibitemShut {NoStop}%
\bibitem [{\citenamefont {Kells}\ \emph {et~al.}(2009)\citenamefont {Kells},
  \citenamefont {Moran},\ and\ \citenamefont {Vala}}]{kells2009}%
  \BibitemOpen
  \bibfield  {author} {\bibinfo {author} {\bibfnamefont {G.}~\bibnamefont
  {Kells}}, \bibinfo {author} {\bibfnamefont {N.}~\bibnamefont {Moran}}, \ and\
  \bibinfo {author} {\bibfnamefont {J.}~\bibnamefont {Vala}},\ }\href {\doibase
  10.1088/1742-5468/2009/03/P03006} {\bibfield  {journal} {\bibinfo  {journal}
  {J. Stat. Mech. Theor. Exp.}\ }\textbf {\bibinfo {volume} {2009}},\ \bibinfo
  {pages} {P03006} (\bibinfo {year} {2009})}\BibitemShut {NoStop}%
\bibitem [{Note2()}]{Note2}%
  \BibitemOpen
  \bibinfo {note} {More precisely, the transverse-field toric code can be
  further mapped via the Xu-Moore model for the $p + ip$ superconductor
  arrays~\cite {PhysRevLett.93.047003} to a square-lattice compass model~\cite
  {PhysRevB.71.195120}, in which the first-order transition line terminating at
  a finite-$T$ critical point was confirmed by using quantum Monte Carlo
  simulations ~\cite {PhysRevLett.98.256402,PhysRevB.78.064402}.}\BibitemShut
  {Stop}%
\bibitem [{\citenamefont {Rau}\ \emph {et~al.}(2014)\citenamefont {Rau},
  \citenamefont {Lee},\ and\ \citenamefont {Kee}}]{PhysRevLett.112.077204}%
  \BibitemOpen
  \bibfield  {author} {\bibinfo {author} {\bibfnamefont {J.~G.}\ \bibnamefont
  {Rau}}, \bibinfo {author} {\bibfnamefont {E.~K.-H.}\ \bibnamefont {Lee}}, \
  and\ \bibinfo {author} {\bibfnamefont {H.-Y.}\ \bibnamefont {Kee}},\ }\href
  {\doibase 10.1103/PhysRevLett.112.077204} {\bibfield  {journal} {\bibinfo
  {journal} {Phys. Rev. Lett.}\ }\textbf {\bibinfo {volume} {112}},\ \bibinfo
  {pages} {077204} (\bibinfo {year} {2014})}\BibitemShut {NoStop}%
\bibitem [{\citenamefont {Yamaji}\ \emph {et~al.}(shed)\citenamefont {Yamaji},
  \citenamefont {Misawa}, \citenamefont {Todo}, \citenamefont {Yoshimi},
  \citenamefont {Kawamura},\ and\ \citenamefont {Kawashima}}]{HPhipre}%
  \BibitemOpen
  \bibfield  {author} {\bibinfo {author} {\bibfnamefont {Y.}~\bibnamefont
  {Yamaji}}, \bibinfo {author} {\bibfnamefont {T.}~\bibnamefont {Misawa}},
  \bibinfo {author} {\bibfnamefont {S.}~\bibnamefont {Todo}}, \bibinfo {author}
  {\bibfnamefont {K.}~\bibnamefont {Yoshimi}}, \bibinfo {author} {\bibfnamefont
  {M.}~\bibnamefont {Kawamura}}, \ and\ \bibinfo {author} {\bibfnamefont
  {N.}~\bibnamefont {Kawashima}},\ }\href {https://github.com/QLMS/HPhi}
  {\bibfield  {journal} {\bibinfo  {journal} {https://github.com/QLMS/HPhi}\ }
  (\bibinfo {year} {unpublished})}\BibitemShut {NoStop}%
\bibitem [{\citenamefont {Rieger}\ and\ \citenamefont
  {Kawashima}(1999)}]{Rieger1999}%
  \BibitemOpen
  \bibfield  {author} {\bibinfo {author} {\bibfnamefont {H.}~\bibnamefont
  {Rieger}}\ and\ \bibinfo {author} {\bibfnamefont {N.}~\bibnamefont
  {Kawashima}},\ }\href {\doibase 10.1007/s100510050761} {\bibfield  {journal}
  {\bibinfo  {journal} {Eur. Phys. J. B}\ }\textbf {\bibinfo {volume} {9}},\
  \bibinfo {pages} {233} (\bibinfo {year} {1999})}\BibitemShut {NoStop}%
\bibitem [{\citenamefont {Xu}\ and\ \citenamefont
  {Moore}(2004)}]{PhysRevLett.93.047003}%
  \BibitemOpen
  \bibfield  {author} {\bibinfo {author} {\bibfnamefont {C.}~\bibnamefont
  {Xu}}\ and\ \bibinfo {author} {\bibfnamefont {J.~E.}\ \bibnamefont {Moore}},\
  }\href {\doibase 10.1103/PhysRevLett.93.047003} {\bibfield  {journal}
  {\bibinfo  {journal} {Phys. Rev. Lett.}\ }\textbf {\bibinfo {volume} {93}},\
  \bibinfo {pages} {047003} (\bibinfo {year} {2004})}\BibitemShut {NoStop}%
\bibitem [{\citenamefont {Nussinov}\ and\ \citenamefont
  {Fradkin}(2005)}]{PhysRevB.71.195120}%
  \BibitemOpen
  \bibfield  {author} {\bibinfo {author} {\bibfnamefont {Z.}~\bibnamefont
  {Nussinov}}\ and\ \bibinfo {author} {\bibfnamefont {E.}~\bibnamefont
  {Fradkin}},\ }\href {\doibase 10.1103/PhysRevB.71.195120} {\bibfield
  {journal} {\bibinfo  {journal} {Phys. Rev. B}\ }\textbf {\bibinfo {volume}
  {71}},\ \bibinfo {pages} {195120} (\bibinfo {year} {2005})}\BibitemShut
  {NoStop}%
\bibitem [{\citenamefont {Tanaka}\ and\ \citenamefont
  {Ishihara}(2007)}]{PhysRevLett.98.256402}%
  \BibitemOpen
  \bibfield  {author} {\bibinfo {author} {\bibfnamefont {T.}~\bibnamefont
  {Tanaka}}\ and\ \bibinfo {author} {\bibfnamefont {S.}~\bibnamefont
  {Ishihara}},\ }\href {\doibase 10.1103/PhysRevLett.98.256402} {\bibfield
  {journal} {\bibinfo  {journal} {Phys. Rev. Lett.}\ }\textbf {\bibinfo
  {volume} {98}},\ \bibinfo {pages} {256402} (\bibinfo {year}
  {2007})}\BibitemShut {NoStop}%
\bibitem [{\citenamefont {Wenzel}\ and\ \citenamefont
  {Janke}(2008)}]{PhysRevB.78.064402}%
  \BibitemOpen
  \bibfield  {author} {\bibinfo {author} {\bibfnamefont {S.}~\bibnamefont
  {Wenzel}}\ and\ \bibinfo {author} {\bibfnamefont {W.}~\bibnamefont {Janke}},\
  }\href {\doibase 10.1103/PhysRevB.78.064402} {\bibfield  {journal} {\bibinfo
  {journal} {Phys. Rev. B}\ }\textbf {\bibinfo {volume} {78}},\ \bibinfo
  {pages} {064402} (\bibinfo {year} {2008})}\BibitemShut {NoStop}%
\end{thebibliography}%
